\documentclass[preprint]{aastex}
\shorttitle{Warps in Triaxial Halos}
\shortauthors{JEON, KIM \& ANN}

\def\spose#1{\hbox to 0pt{#1\hss}}
\newcommand\lsim{\mathrel{\spose{\lower 3.0pt\hbox{$\mathchar"218$}}
     \raise 2.0pt\hbox{$\mathchar"13C$}}}
\newcommand\gsim{\mathrel{\spose{\lower 3.0pt\hbox{$\mathchar"218$}}
     \raise 2.0pt\hbox{$\mathchar"13E$}}}
\newcommand\msun{{\rm \,M_\odot}}

\begin{document}
\title{Galactic Warps in Triaxial Halos}
\author{Myoungwon Jeon,\altaffilmark{1} Sungsoo S. Kim,\altaffilmark{1,3} and
Hong Bae Ann\altaffilmark{2}}
\altaffiltext{1}{Dept. of Astronomy \& Space Science, Kyung Hee University,
Yongin-shi, Kyungki-do 449-701, Korea; myjeon@ap4.khu.ac.kr,
sungsoo.kim@khu.ac.kr}
\altaffiltext{2}{Division of Science Education, Pusan National University,
Pusan 609-735, Korea; hbann@pusan.ac.kr}
\altaffiltext{3}{Corresponding author}

\begin{abstract}
We study the behaviors of galactic disks in triaxial halos both numerically
and analytically to see if warps can be excited and sustained in triaxial
potentials.  We consider the following two scenarios: 1) galactic disks
that are initially tilted relative to the equatorial plane of the halo
(for a pedagogical purpose),
and 2) tilted infall of dark matter relative to the equatorial plane of the
disk and the halo.  With numerical simulations of 100,000 disk particles in a
fixed halo potential, we find that in triaxial halos, warps can be excited
and sustained just as in spherical or axisymmetric halos but they show some
oscillatory behaviors and even can be transformed to a polar-ring
system if the halo has a prolate-like triaxiality.  The non-axisymmetric
component of the halo causes the disk to nutate, and the differential
nutation between the inner and outer parts of the disk generally makes
the magnitude of the warp slightly diminish and fluctuate.
We also find that warps are relatively weaker in oblate and oblate-like
triaxial halos, and since these halos are the halo configurations of disk
galaxies inferred by cosmological simulations, our results are consistent
with the fact that most of the observed warps are quite weak.
We derive approximate formulae for the torques exerted on the disk by the
triaxial halo and the dark matter torus, and with these formulae we
successfully describe the behaviors of the disks in our simulations.
The techniques used in deriving these formulae could be applied for
realistic halos with more complex structures.
\end{abstract}

\section{INTRODUCTION}
\label{sec:intro}

The outer parts of the disks in most spiral galaxies, even in isolated ones,
are warped (Ann \& Park 2006 and references therein),
and this implies that the warps must be either long-lived or be repeatedly
excited.  Early proposals for the excitation mechanism of disk warps
include the intergalactic gas flow past the disk (Kahn \& Woltjer 1959),
tidal distortion due to the Magellanic Clouds (Elwert \& Hablick 1965),
and the free mode of oscillation of a disk (Lynden-Bell 1965).

Extending the work by Lynden-Bell, Hunter and Toomre (1969) studied
the warp in terms of discrete modes of oscillation of an isolated, thin,
self-gravitating disk and found that a long-lived warp can correspond to
a simple discrete mode only when the disk has an unrealistically sharp edge.
Later, Sparke and Casertano (1988) showed that self-gravitating disks
even with realistic density profiles may have long-lived discrete modes
if the disk is embedded in the fixed potential of an axisymmetric halo
(such modes are called the modified-tilt mode).

Dekel \& Shlosman (1983) and Toomre (1983) suggested a misalignment between
the disk and the non-spherical halo as the cause of the warp, and Dubinski
\& Kuijken (1995) and Ideta et al. (2000), among others, performed simulations
for such configuration.  With simulations of tilted disks in fixed,
axisymmetric halos, Ideta et al. found that prolate halos could sustain
galactic warps while the warping in oblate halos continues to wind up and
disappears fairly quickly.

However, dynamical friction between the disk and halo may cause the warp
to disperse within timescales much shorter than a Hubble time (Nelson and
Tremaine 1995).  Numerical studies employing {\it live} halos (Dubinski \&
Kuijken 1995; Binney, Jiang, \& Dutta 1998), which can properly consider
the reaction of the halo to the disk, confirmed that dynamical friction may
play an important role in the disappearance of galactic warps on a time
scale shorter than a Hubble time.  Thus a successful scenario for the warp
should not give rise to significant dynamical friction.  Plausible scenarios
that do not suffer the damping problem
include the reorientation of a massive galactic halo by the cosmic infall
(Ostriker \& Binney 1989; Jiang \& Binney 1999; Shen \& Sellwood 2006)
and the external torque produced in the disk by the accretion of the
intergalactic medium (Revaz \& Pfenninger 2001; Lopez-Corredoira et al. 2002).

Shen \& Sellwood (2006) performed simulations of an idealized form of cosmic
infall on to a disk galaxy and obtained warps that persist for a few Gyr.
They found that the damping from the live halo is very weak compared to
that in initially tilted disk scenarios, and attributed the difference
to the facts that 1) the source of the torque in their model resides in the
outer halo only, 2) the precession rate of their disk is very low, and
3) the effect of the torque in their model kicks in gradually.

It appears that the model by Shen \& Sellwood (2006) is a highly feasible
way of exciting and sustaining a warp, because their warps closely resemble
those observed and do not significantly suffer the damping problem.
But their halo is modeled to be spherical, wherease many numerical studies
on the shape of the dark matter halos anticipate that the triaxiality of
the dark matter halos is not negligible (Jing \& Suto 2002; Bailin \&
Steinmetz 2005; Allgood et al. 2007, among others).  Thus it is worth
checking if the results of the cosmic infall scenario are significantly
altered when the halo is modeled to be axisymmetric or triaxial.
Furthermore, as the triaxiality of the halo exerts a torque that
periodically increases and decreases the inclination angle of the tilted
disk (see \S \ref{sec:model}), the initially tilted disk scenario deserves
a revisit to see if the extra torque by the halo triaxiality can diminish
the effect of the damping.

For these reasons, in the present paper we study the evolution of the
self-gravitating disks in axisymmetric and triaxial halos for
two scenarios: 1) initially tilted disks in triaxial halos and
2) tilted cosmic infall in axisymmetric and triaxial halos.
As discussed above, the first scenario does not appear to be a plausible
mechanism for real warps, but we first study this scenario in detail
to understand the influence of the triaxial potential alone to the disk.
After successfully describing it with simulations and some analytical
formulae, we then analyze the behaviour of the warp in the second scenario.

As a first step, the halos in the present paper are fixed, and this will
enable us to concentrate on studying how the axisymmetry and/or triaxiaility
of the halo alters the evolution of the warp compared to the spherical or
axisymmetric cases.
As a subsequent study to the present one, we will examine the disks
in ``live'' axisymmetric and triaxial halos, which can properly treat
the effect of dynamical friction between the disk and the halo.

Note that the aim of the present study is to understand the fundamental
dynamics between the disk and the axisymmetric or triaxial halo for the
above two scenarios, rather than to find the conditions that best reproduce
the various observed characteristics of the warp.  Still, we will briefly
compare our simulations and observed warps in \S \ref{sec:summary}.

Warps are generally more extended in radio observations of gas than
in optical observations of stars.  However, in the present study, we only
consider the stellar component of the disk to avoid the model complexity
and computing cost involved with the consideration of the hydrodynamic (and
even magnetohydrodynamic) effects.  Thus the warps in our simulations are to
be compared to the inner parts of the observed, extended warps.

We describe the models and the numerical method in \S \ref{sec:model},
and discuss some theoretical backgrounds necessary for the analyses
of our simulations in \S \ref{sec:theory}.  The results and analyses of our
simulations for initially tilted disks are presented in \S \ref{sec:tilted},
and those for the tilted cosmic infall scenario are presented in
\S \ref{sec:infall}.  We summarize and discuss our results in
\S \ref{sec:summary}.

\section{MODELS}
\label{sec:model}

For the simulations presented here, we use a parallel version of the tree
N-body code named GADGET (Springel, Yoshida, and White 2001).
The code can calculate hydrodynamic forces using the smoothed particle
hydrodynamics (SPH) technique, but we only consider gravitational forces
for the present work, i.e. we do not consider the effects of gas component
in the disk.  Gadget was the choice of numerical method also for a recent
study on tidal structures in a disk galaxy created by gravitational
interactions with a perturbing companion (Oh et al. 2008).

Our model galaxies consist of a disk and a dark matter halo.  The disk
is represented by 100,000 particles, and the halo is modeled by
a fixed potential.  Thus the evolution of the disk is governed by
the self-gravity within the disk and the external gravity from the
halo (plus the dark matter torus for the cosmic infall scenario),
but the effect of the disk on the halo and its reaction back to
the disk, such as dynamical friction, is not considered.  In this way,
we would be able to observe the role of various triaxial halos on the
excitation and maintenance of the warp separately from the effect of
mutual back reaction.

Following Ideta et al. (2000), we adopt the compound galaxy model
by Hernquist (1993) and modify it for triaxial halos.  The distribution
of the disk particles is given by
\begin{equation}
\label{rho_d}
	\rho_d (R,z) = \frac{M_d}{4 \pi R_d^2 z_d} \exp (-R/R_d) {\rm sech}^2
                       \left ( \frac{z}{z_d} \right ),
\end{equation}
where $R$ is the galactocentric radius projected on to the galactic plane,
$z$ the vertical height from the plane, $M_d$ the disk mass, $R_d$ the
radial scale length, and $z_d$ the vertical scale thickness.  The density
profile for the triaxial halo is given by
\begin{equation}
\label{rho_h}
	\rho_h (\mu) = \frac{M_h}{2 \pi a b c} \frac{1}{\mu (1+\mu)^3 },
\end{equation}
where $M_h$ is the halo mass and $\mu$ is defined by
\begin{equation}
\label{mu}
        \mu^2 = \frac{x^2}{a^2} + \frac{y^2}{b^2} + \frac{z^2}{c^2}.
\end{equation}
The triaxiality parameters $a$, $b$, and $c$ are the scale lengths
along the $x$, $y$, and $z$ axes.  Our triaxial halos are named either
``oblate-like'' or ``prolate-like'': Oblate-like halos have major
and intermediate axes in the equatorial plane, while prolate-like halos
have minor and intermediate axes in the equatorial plane.
Table \ref{table:simul} lists $a$, $b$, and $c$ values of our models.

Using the method of ellipsoidal shells (Kellogg 1953; Chandrasekhar 1969;
Binney \& Tremaine 2008), the halo potential can be written as
\begin{equation}
\label{Phi_h}
	\Phi_h (\mu) = -\frac{GM_h}{2} \int_0^\infty\frac{{\rm d}u}
			{\sqrt{a^2+u}\sqrt{b^2+u}\sqrt{c^2+u}[1+\mu(u)]^2},
\end{equation}
where
\begin{equation}
\label{mu_u}
        \mu^2(u) = \frac{x^2}{a^2+u} + \frac{y^2}{b^2+u} + \frac{z^2}{c^2+u},
\end{equation}
and we calculate the force exerted by the halo by numerically differentiating
this potential.\footnote{After performing all of our simulations, we realized
that Merritt \& Fridman (1996) present an integral form of the force for our
halo potential.  They give the triaxial generalization of Dehnen (1993)
models (both potentials and forces), and our halo model is a $\gamma=1$ case
of the Dehnen model family.}

Our simulations are in units of $G=1$, $R_d=1$, and $M_d=1$, where $G$ is
the gravitational constant.  If these units are scaled to physical values
appropriate for the Milky Way, i.e. $R_d=3.5$~kpc and $M_d=5.6 \times 10^{10}
\msun$, unit time and velocity become $1.31 \times 10^7$~yr and 262~km~s$^{-1}$.
We set $z_d=0.2$ and $M_h=17.8$ for all of our simulations.  These numbers
result in the minimum Toomre $Q$ parameter of about 1.5,
thus the disk is stable against the bar instability.
The orbital period at the half-mass radius, $R \simeq 1.7$, is 13.4 in
our system of units.  We perform our simulations until $T=600$, which is
$\sim 45$ orbital periods at the half-mass radius of the disk.
To achieve an equilibrium configuration, we sampled velocities from Gaussian
distributions with means and dispersions derived from the Jeans equation
following the prescription by Hernquist (1993).
The total energy is conserved to better than 0.2 per cent.

In our first set of simulations (initially tilted disk models), the disks
are initially tilted by $30^\circ$
with respect to the equatorial ($x$-$y$) plane of the halo in order to
initiate the vertical oscillation of the disk particles.

In the second set of simulations, the disks are initially in the equatorial
plane of the halo with no systematic vertical motions, and following Shen
\& Sellwood (2006), we assume that the late infallng material in hierarchical
galaxy formation models forms a gradually growing torus of radius
$R_t=10.7 R_d$.  Its mass increases linearly from 0 at $T=0$ to $M_t=2.5 M_d$
at $T=200$, and its cross-section has a Gaussian density profile
with a standard deviation of $0.2 R_t$.  The torus plane is
inclined at $15 \degr$ relative to the equatorial plane of the halo, and
its the ascending node aligns with the negative $y$-axis.  The torus is
represented by 5,000 particles that initially have circular
motions with velocities equal to the local circular speed.  We assume
that the torus does not precess as its precession rate due to the
interaction with the disk or halo is much longer than our simulation
period and also assume that the torus is dynamically stable for the sake
of simplicity.  To avoid the instability of a cold, self-gravitating torus,
we keep the torus particles at their initial positions during the whole
simulation period.  Note that our models for the cosmic infall scenario
are different from Shen \& Sellwood (2006) in that their halo density
profile is based on the King model and their torus is located at
a farther distance compared to the size of the disk ($R_t=15 R_d$).
The latter implies that the self-gravity within the disk will be relatively
less important in our models.

\section{THEORETICAL BASIS}
\label{sec:theory}

Following Ideta et al. (2000), we analyze the behavior of the disk in terms
of the relative differences in the inclination and in the line of ascending
node (LON) between the inner and outer parts of the disk.  In order for the
disk to be observed as a warp, the inclination of the outer disk needs to
be different enough from that of the inner disk, but the LONs of both inner
and outer disks should not wind up and need to nearly coincide.

\begin{figure}
\epsscale{0.8}
\plotone{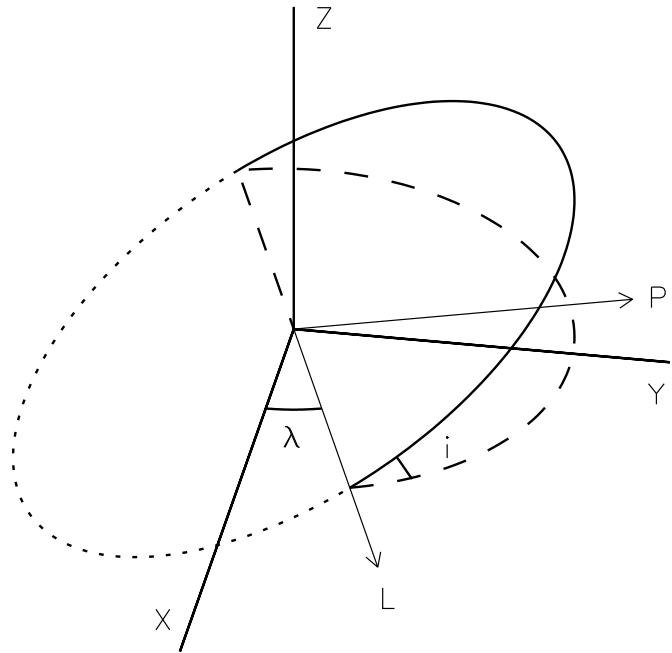}
\caption
{\label{fig:ring}Definitions of $L$- and $P$- axes.  The positive $L$-axis
coincides with the line of ascending node (LON) of the ring, and the $P$-axis
is normal to the $L$-axis in the halo equatorial plane such that $\vec L
\times \vec P$ aligns with the positive $z$-axis.  Note that both $L$- and
$P$- axes always lie in the $x$-$y$ plane.}
\end{figure}

The disk precesses when the torque applied to the disk has a non-zero LON
component, and the precession makes the longitude of the LON, $\lambda$,
evolves.  The inclination of the disk, $i$, evolves
when the torque has a non-zero component that is normal to both the LON
and the $z$-axis.
Since the components of the forementioned torques are either parallel or normal
to the LON, we adopt a rotating frame where the $L$-axis aligns with the
LON of the disk and the $P$-axis is normal to the $L$-axis
in the halo equatorial plane such that when the positive $L$-axis aligns
with the positive $x$-axis, the positive $P$-axis aligns with the positive
$y$-axis (see Fig. \ref{fig:ring}).
In this frame, the LON evolves due to the $L$-component of the
torque, $T_L$ (see eq. [\ref{rate2}]), while the inclination evolves due to
the $P$-component of the torque, $T_P$ (see eq. [\ref{rate1}]).

Our disks have angular momenta whose $z$-component values are positive,
i.e. our disks rotate counterclockwise when viewed from the positive
$z$-axis (the $V_c$ values in eqs. [\ref{rate1}] and [\ref{rate2}]
are defined to be positive for such rotations).  Thus a positive (negative)
$T_L$ will induce a prograde (retrograde) precession, while a positive
(negative) $T_P$ will result in the decrease (increase) of the inclination.

\subsection{Torques by the Triaxial Halo}

Approximate formulae for the torques by triaxial halos are derived in
Appendix \ref{appen:halo}.  The characteristics of the halos are described
with the coefficients $J_{20}$ and $J_{22}$ (see eqs. [\ref{coef20}] and
[\ref{coef22}] for their definitions, and Table \ref{table:jvalues} for
their values at $r=2$ and 4; $r$ is a 3-dimensional galactocentric radius),
which are related to the spherical harmonic
functions of degree $l=2$ and order $m=0$,2.
Oblate halos have negative $J_{20}$ values and prolate
halos have positive $J_{20}$.  While both oblate and prolate halos have
$J_{22}=0$, triaxial halos have a non-zero $J_{22}$.  Thus equations
(\ref{torque1}) and (\ref{torque2}) imply that axisymmetric halos exert
$T_L$ only, whereas triaxial halos exert both $T_L$ and $T_P$.
As a result, axisymmetric halos can cause changes in $\lambda$ only,
but triaxial halos can cause changes in both $i$ and $\lambda$ (see eqs.
[\ref{rate1}] and [\ref{rate2}]).

Another consequence of the triaxial halo is the oscillations in the
magnitudes of the torques.
Equations (\ref{torque1}) and (\ref{torque2}) show that the
contribution of the triaxial component $J_{22}$ to both $T_L$ and $T_P$ has
a sinusoidal form ($\cos 2 \lambda$ or $\sin 2 \lambda$).  In this subsection,
we discuss the behavior of the disk in triaxial halos with $a>b$ (all of our
triaxial halos except those in models CI-OL7b and CI-PL7b have $a>b$), which
have positive $J_{22}$ values.
In oblate-like halos, the LON undergoes a retrograde precession (because of
the negative $J_{20}$), thus the absolute value of $T_L$ decreases
(increases) when $\lambda$ is in the first and third (second and fourth)
quadrants.  In prolate-like halos, the LON advances forward (because of
the positive $J_{20}$) and thus as in the oblate-like halos, the absolute
value of $T_L$ decreases (increases) when $\lambda$ is in the first and
third (second and fourth) quadrants.  These effects cause the precession
rate to oscillate twice during one full precession of the LON
($|d\lambda/dt|$ decreases [increases] when $\lambda$ is in the first
and third [second and fourth] quadrants).

As for $T_P$, the sign of its value is more important than the decrease or
increase of the value.  In both oblate-like and prolate-like halos,
$T_P$ will be negative (positive) in the first and third (second and fourth)
quadrants.  This causes $i$ to oscillate twice during one full precession
of the LON ($di/dt$ is positive [negative] in the fist and third [second and
fourth] quadrants), i.e., the disk nutates as it precesses.

\subsection{Torques by the Accreting Torus}

The accreting torus is inclined at a small angle ($15 \degr$) relative
to the halo equatorial plane, thus the torus has a negative $J_{20}$,
i.e., the most significant non-spherical component of the torus is the
oblateness.  Furthermore, since the mass distribution of the torus
has a sinusoidal deviation from the equatorial plane with an azimuthal
period of $2 \pi$, the torus has a non-zero $J_{21}$ component (see Table
\ref{table:jvalues} for these values at $r=2$ and 4), which
is not present in triaxial halos and is much more important
than the $J_{22}$ component of the torus.

Equations (\ref{rate1t}) and (\ref{rate2t}) show that $di/dt$ and
$d\lambda/dt$ are sinusoidal functions of $\lambda$ and imply that $i$
and $\lambda$ will oscillate around certain equilibrium values.
The circular velocity of the disk is nearly constant in the outer disk
region, thus equations (\ref{rate1t}), (\ref{rate2t}), and (\ref{coef20t})
through (\ref{coef22t}) imply that $di/dt$ and $d\lambda /dt$ by the torus
are approximately proportional to $r$, whereas $di/dt$ and $d\lambda /dt$
by the halo decrease as $r$ increases.

\subsection{Torques by the Inner Disk}
\label{subsec:theory_id}

We adopt a simple two-disk (inner and outer disks) model to describe the
effect of the self-gravity within the disk.  
Equations (\ref{rate1d}) and (\ref{coef21d}) imply that if $\lambda$ of
the outer disk is ahead (in the sense of disk spin) of that of the inner
disk (i.e., a leading spiral of the LON), $di/dt$ is positive, and vice
versa.  Equations (\ref{rate2d}), (\ref{coef20d}) and (\ref{coef21d}) state
that $d\lambda /dt$ by the inner disk is negative on average when the
inclination of the inner disk, $i_{id}$, is smaller than $\sim 55 \degr$,
and has a large $\lambda$ dependence when $i$ of the outer disk is small.

\section{INITIALLY TILTED DISKS}
\label{sec:tilted}

Here we present the results obtained from our simulations for the
initially tilted disks and analyze them by comparing the axisymmetric
and triaxial cases.

\subsection{Oblate vs. Oblate-like Halos}
\label{subsec:td-o}

\begin{figure}
\epsscale{1.1}
\plotone{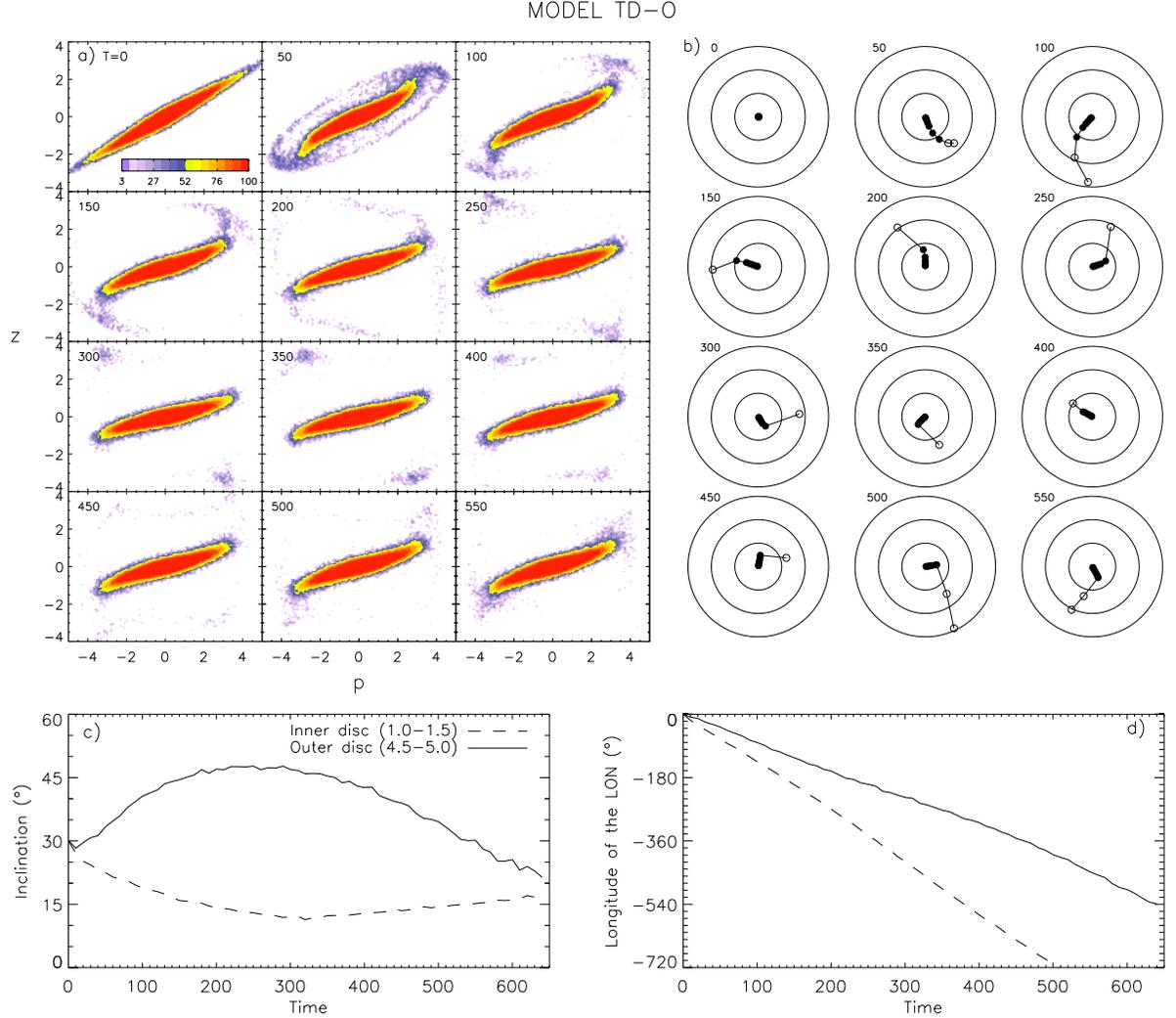}
\caption
{\label{fig:td-o}Disk Evolution of model TD-O.  a) Projected density
distribution of the simulation particles as seen from the ascending line of
node (LON) of the disk.  The abscissa aligns with the $x$-$y$ plane.
The color bar shows the level of the surface density in units of \# particles
per projected area of $0.1^2$.
b) $i_{rel}$-$\lambda$ plot (the `Briggs diagram') for 9 equal-size radial
bins between $r=1.0$ and 4.6 (thus each bin has a width of 0.4).
The distance from the plot center corresponds to $i_{rel}$, the inclination
relative to the central region of the disk, and the azimuthal angle
from the positive $x$-axis corresponds to $\lambda$, the longitude of the
ascending LON.  Each concentric circle marks $10 \degr$.  The inclination of
the disk center is defined to be the average $i$ of the two innermost radial
bins($r=1.0$--1.8).  Open symbols are for the radial bins whose peak
projected density is smaller than 20 particles per $0.1^2$ area.
c) $i$ evolution of two radial bins that represent the inner and outer
disks (the radial ranges of these bins are indicated in the plot).
d) $\lambda$ evolution of the same two radial bins as in panel c).}
\end{figure}

Figure \ref{fig:td-o} shows the evolution of the initially tilted disk
in an oblate halo (model TD-O).  This simulation is set up in the same
way as the oblate model of Ideta et al. (2000) in order to check the
agreement between the two simulation results and to describe the
results of our triaxial models in comparison to the axisymmetric ones.
We find that our model TD-O gives almost the same result as the oblate
model of Ideta et al.  First we briefly discuss the evolution of the disk
in model TD-O in terms of torques following the analysis by Ideta et al.
(2000).

Oblate halos exert negative $T_L$ on the disk, and the oblate halo adopted
here has the absolute precession rate that decreases as $r$ increases.
Thus the LON undergoes a retrograde precession with a shape of a leading
spiral.  Since the LON of the outer disk is ahead (in the sense of disk spin)
of the LON of the inner disk, the inclination of the outer disk increases
forming a type I warp (terminology used by Sparke \& Casertano 1988),
which bends upward away from the halo equatorial plane.  This is because
the inner disk behind the outer disk exerts a negative $T_P$ on to the outer
disk, which results in an increase of $i$ (see \S \ref{subsec:theory_id}).
However, a larger inclination causes a slower precession ($|d\lambda/dt|
\propto \cos i$; see eq. [\ref{rate2}]), thus the LON of the outer disk
leads even further and does not have a chance to catch up the inner disk.
As the LONs of the inner and outer disks become significantly misaligned,
the disk is not observed as a warp anymore.

This can be clearly seen in the $i_{rel}$-$\lambda$ diagrams (so-called
``Briggs diagrams'', Briggs 1990) in panel $b$, which effectively shows
the radial profiles of $\lambda$  and the inclination relative to that at
the central region of the disk, $i_{rel}$.  Each circle
represents an annulus of $0.4 R_d$ between $1 R_d$ and $4.6 R_d$.  The disk
particles in some of the outermost annuli tend to spread out in the projected
sky making them not recognized as a part of the disk in actual observations,
so we use open circles to indicate the annuli that have a peak projected
density less than 20 particles per $0.1^2$ area.
In such diagrams, an observable warp would be manifested as an extended
alignment of filled circles.  The filled circles are extended
out to $\sim 10 \degr$ only until $T \simeq 200$ and remain inside $i_{rel}
\simeq 7\degr$ afterwards.  Panel $a$ indeed shows that a warp
phenomenon is significant only until this epoch.

\begin{figure}
\epsscale{1.1}
\plotone{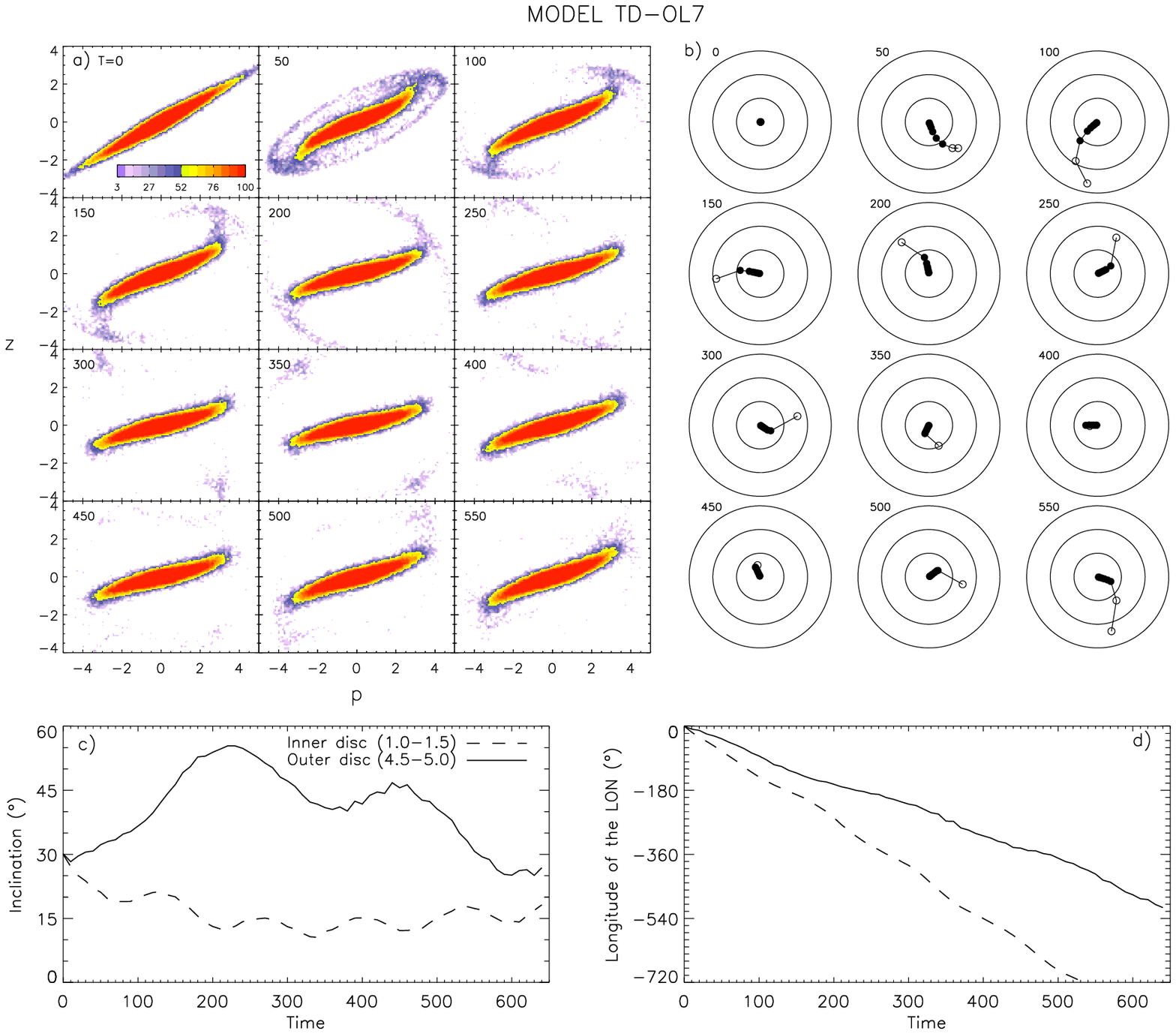}
\caption
{\label{fig:td-ol7}Same as Figure \ref{fig:td-o}, but for model TD-OL7.}
\end{figure}

Figure \ref{fig:td-ol7} plots the evolution of a disk in our moderately
oblate-like halo, model TD-OL7 (note that all of our oblate and oblate-like
halos have similar $[a^2+b^2]/c^2$ ratios
so that they have similar $J_{20}$ values and thus similar precession rates).
It shows that the triaxiality of the halo does not greatly change the
global aspect of the evolution.  The most significant difference between
the oblate and oblate-like cases is that $i$ and $\lambda$ oscillate
periodically in the latter, which is the result of $P$- and $L$-component
torques from the triaxial component of the halo, respectively---$T_P$ and $T_L$
are sinusoidal functions when $J_{22}$ has a non-zero value (see eqs.
[\ref{torque1}] and [\ref{torque2}]).
The phase of the oscillation coincides with the quadrant of the LON,
which indicates that the oscillation is indeed caused by the triaxial
component of the halo.  As discussed in \S~\ref{sec:theory}, when $\lambda$
is in the first or third quadrant, the disk experiences a negative
$P$-component torque and the inclination increases.  Likewise, the
inclination decreases when $\lambda$ is in the second or fourth quadrant.

\begin{figure}
\epsscale{1.1}
\plotone{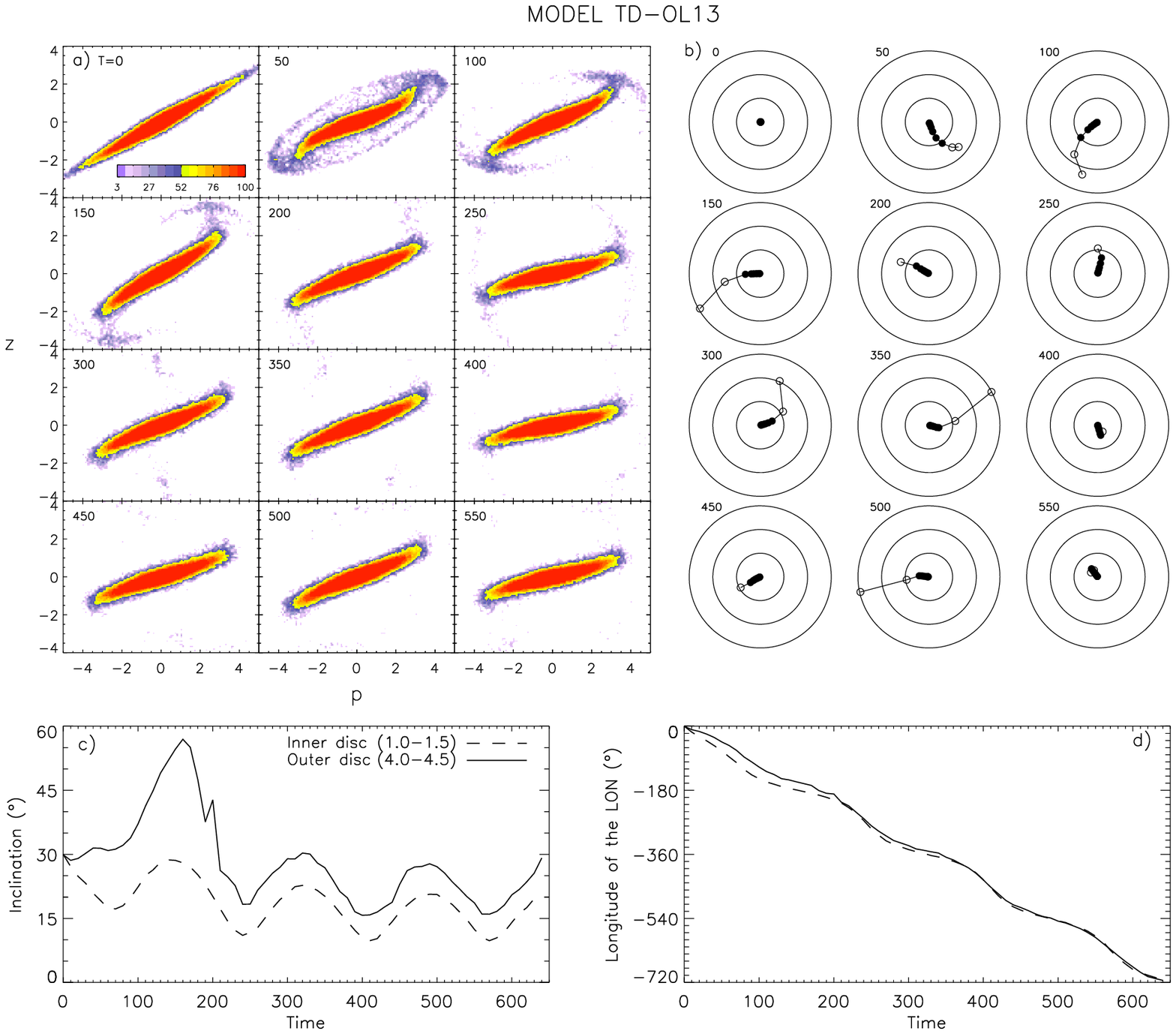}
\caption
{\label{fig:td-ol13}Same as Figure \ref{fig:td-o}, but for model TD-OL13.}
\end{figure}

A similar evolutionary picture is observed even in the more extremely
oblate-like case, model TD-OL13.  Figure \ref{fig:td-ol13} shows that
the global evolution of this model is very similar to that of model TD-OL7.

In both models TD-OL7 and TD-OL13, the magnitude of the warp slightly
oscillates, but the average magnitude of the warp is similar to that
of model TD-O.  To summarize, when $J_{20}$ of the halo is negative,
the warp in an initially tilted disk scenario is very weak except at the
beginning, and even the triaxiality of the halo is not effective in increasing
the magnitude of the warp.

\subsection{Prolate vs. Prolate-like Halos}
\label{subsec:td-p}

\begin{figure}
\epsscale{1.1}
\plotone{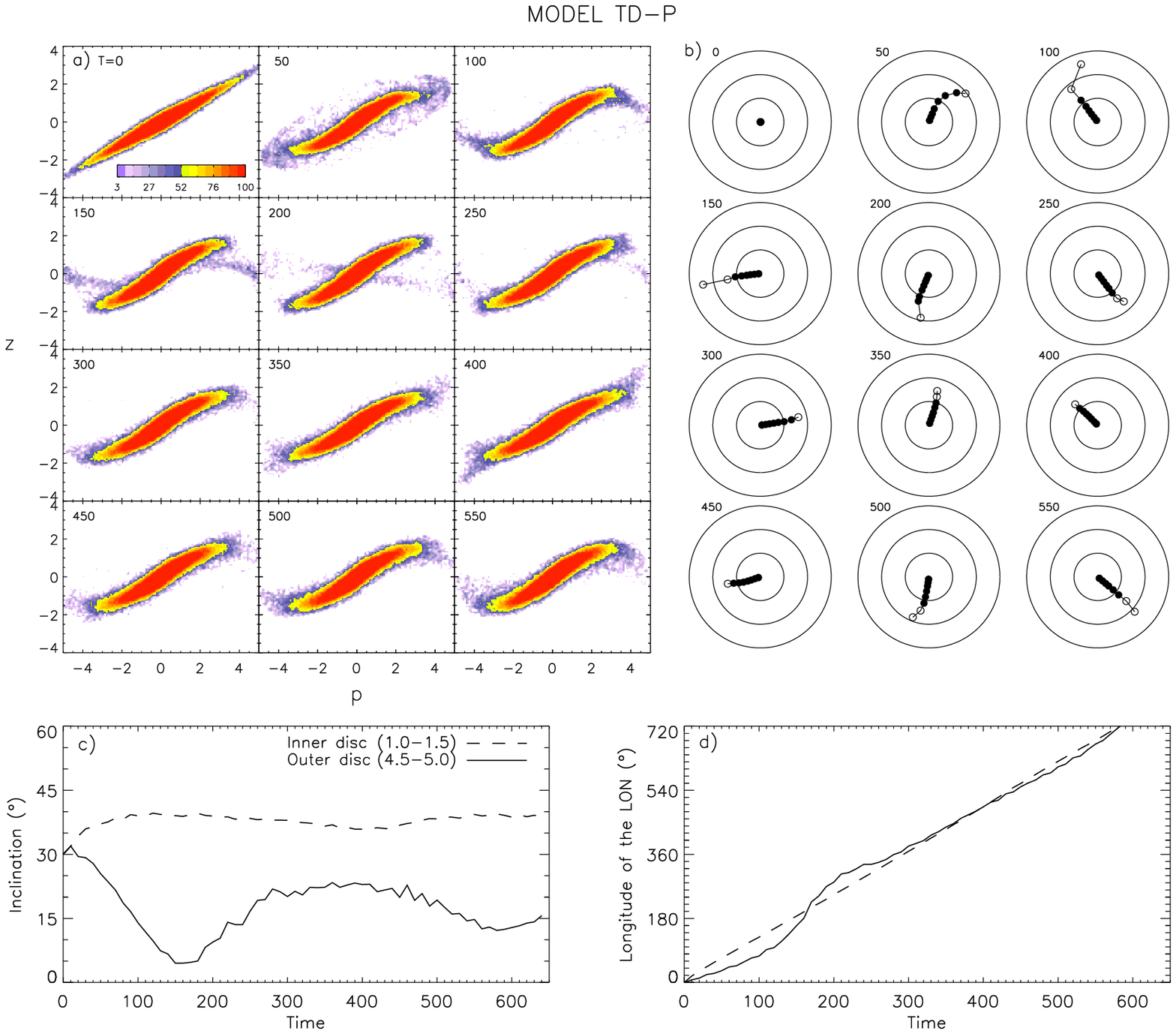}
\caption
{\label{fig:td-p}Same as Figure \ref{fig:td-o}, but for model TD-P.}
\end{figure}

The evolution of the disk in a prolate halo, model TD-P, is shown in
Figure \ref{fig:td-p}.  Prolate halos exert positive $T_L$ on the disk,
and our prolate halos have an absolute precession rate that decreases
as $R$ increases.  Thus the LON undergoes a prograde precession with
a shape of a trailing spiral.  Because the LON of the outer disk is behind
(in the sense of disk spin) that of the inner disk, $T_P$ exerted by the
inner disk on to the outer disk is positive, and the inclination of the
outer disk decreases (see \S \ref{subsec:theory_id})
forming a type II warp, which bends down toward the halo equatorial plane.
Since a smaller inclination causes a faster precession ($|d\lambda/dt|
\propto \cos i$; see eq. [\ref{rate2}]),
the LON of the outer disk catches up that of the inner disk.
If the LON of the outer disk outruns that of the inner disk, the outer
disk will have a larger inclination and its precession rate will become
smaller.  Thus the self-gravity between the inner and outer disks
plays a role of restoring force that leads to an alignment of the LONs between
the inner and outer disks.  The simulation shows that after about one
revolution of the precession, the inner and outer disks
settle into a stable equilibrium, and the LONs of the two disks remain
close to each other afterwards, making the disk observed as a warp.

\begin{figure}
\epsscale{1.1}
\plotone{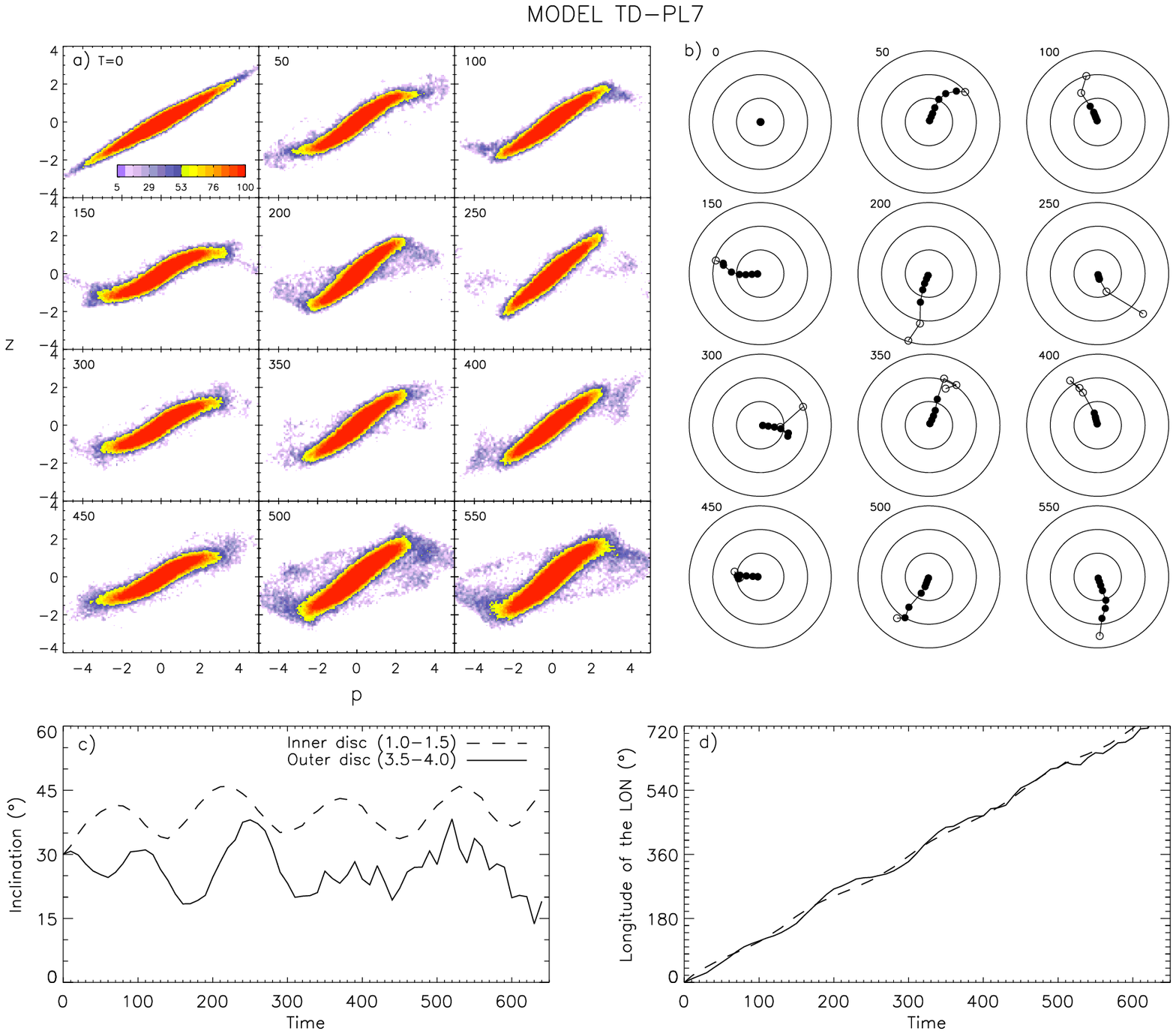}
\caption
{\label{fig:td-pl7}Same as Figure \ref{fig:td-o}, but for model TD-PL7.}
\end{figure}

This analysis can be generally applied to the evolution of the disk in
a moderately prolate-like halo, model TD-PL7 (Fig. \ref{fig:td-pl7}).
Note that all of our prolate and prolate-like halos have similar
$(a^2+b^2)/c^2$ ratios so that they have similar precession rates.
Again, the most considerable difference from the prolate case is the
oscillation in both $i$ and $\lambda$.
Since the LON of the inner disk advances faster than
that of the outer disk in the beginning of the simulation, the oscillation
phase of the inner disk is ahead of that of the outer disk afterwards.
Because of this phase lag, the inclination gap between the two disks
increases and decreases periodically (without the phase lag, the inclinations
of the two disks would oscillate with an almost constant gap).  Therefore,
as seen in panels $a$ and $b$, the magnitude of the warp oscillates with
a period of precession as well.  The warp nearly disappears at
$T \simeq 250$ and 400, and grows back to its peak at $T \simeq 350$
and 500.  Note that when the magnitude of the warp reaches its peak, it
becomes larger than that of the prolate case with a similar prolateness
($c^2/[a^2+b^2]$).  The $i_{rel}$ value of the outermost annulus
remains inside $15 \degr$ in model TD-P, but it reaches near
$20 \degr$ at $T=350$ and 500 in model TD-PL7.  Thus the
triaxiality of the prolate-like halo periodically suppresses and
augments the magnitude of the warp.

\begin{figure}
\epsscale{1.1}
\plotone{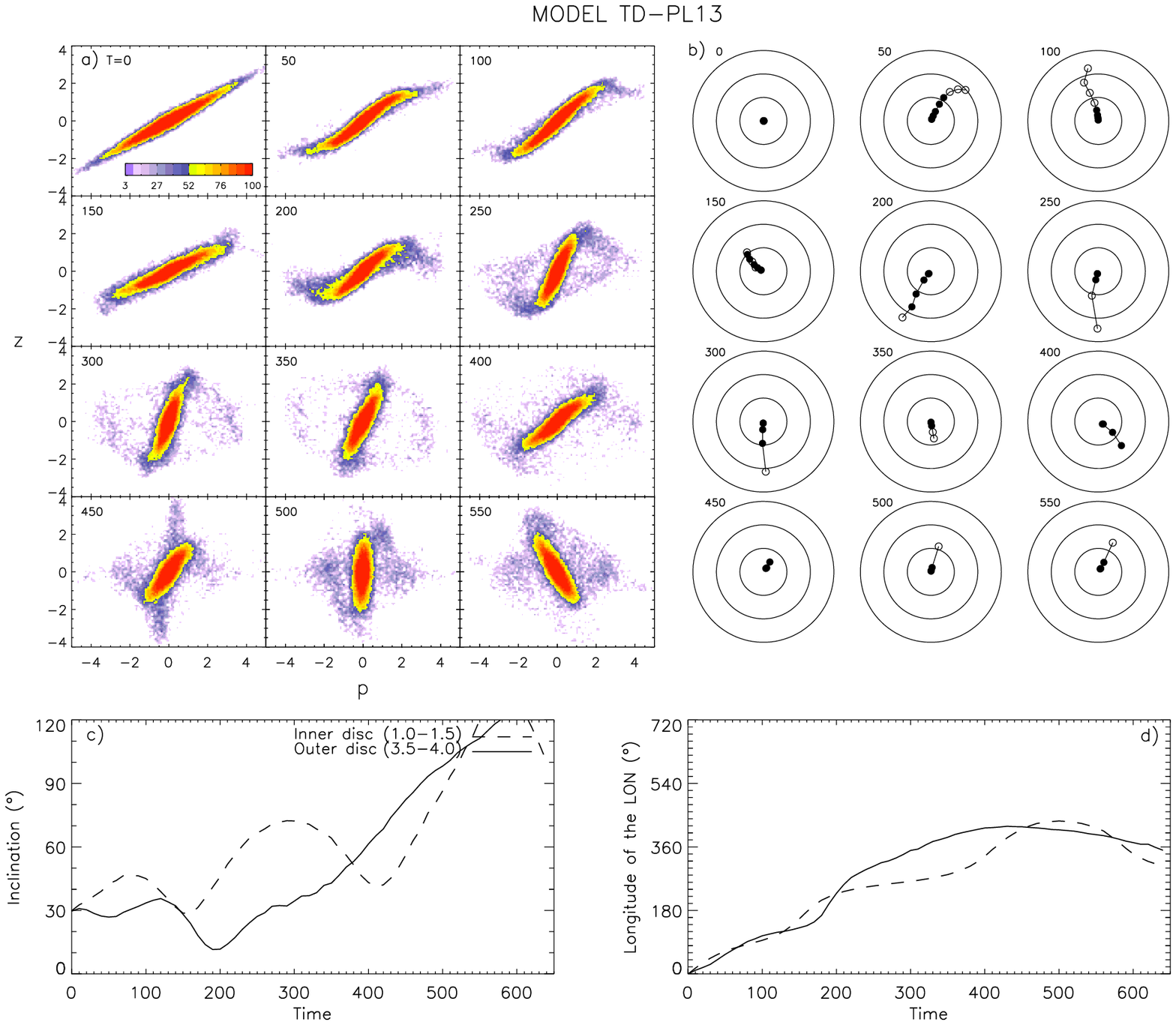}
\caption
{\label{fig:td-pl13}Same as Figure \ref{fig:td-o}, but for model TD-PL13.}
\end{figure}

In a more extremely prolate-like halo (model TD-PL13), the disk evolves
quite differently from the above two cases.  As seen in panel $c$ of
Figure \ref{fig:td-pl13}, the inclinations of both inner and outer disks
increase fairly quickly while oscillating.  This appears to be because
of the $\sin i$ factor in $di/dt$ term (eq. [\ref{rate1}])--When the
the magnitude of the triaxiality ($J_{22}$) is large, the peak of the
oscillating $i$ can reach higher, and at a higher $i$, $di/dt$ is larger.
Thus when $J_{22}$ is larger than a certain critical value, the peak
$i$ value in each oscillation gradually increases and eventually
$i$ becomes larger than $90 \degr$.  This happens at $T \simeq 500$ for
model TD-PL13, after which the LON evolves in the reverse direction because
the $\cos i$ term in $d \lambda /dt$ (eq. [\ref{rate2}]) becomes negative.

The large $J_{22}$ value and thus the large oscillation in $i$ of model
TD-PL13 causes the outer disk to disperse so severely that it is not
recognized as a part of the disk anymore after $T \sim 450$, making the size
of the whole disk appear to be much smaller than in the beginning.  The debris
from the outer disk spreads out randomly around the inner disk
after this epoch, and the warped configuration nearly disappears.

Although panel $c$ of Figure \ref{fig:td-pl13} shows the evolution up to
$T=650$, we have evolved model TD-PL13 up to $T=800$ and find that
after $T \simeq 400$, $i$ and $\lambda$ of the inner disk keep oscillating
around $90 \degr$ and $360 \degr$, respectively, indicating that the
disk plane tumbles about the minor axis of the halo.  Such a phenomenon
has been also observed in numerical and analytical studies on the
dynamics of stars and gas in triaxial potentials (Aarseth \& Binney 1978;
Steiman-Cameron \& Durisen 1984; Arnaboldi \& Sparke 1994, among others):
the particles in triaxial halos prefer the orbits whose principal plane
tumbles about its minor axis.

To summarize, when $J_{20}$ of the halo is positive, an initially tilted
disk can excite and sustain the warped configuration for a considerable time
as far as the triaxial component of the halo is not too large, and a mild
triaxiality can augment the peak magnitude of the warp.

\section{TILTED COSMIC INFALL}
\label{sec:infall}

Now we present and analyze our simulations for the cosmic infall scenario.
In these simulations, the disk is initially in the equatorial plane of the
halo and the dark matter torus, whose mass gradually increases over $T=100$,
is tilted at $15 \degr$ relative to the halo equatorial plane such that its
ascending LON coincides with the negative $y$-axis.

\subsection{Spherical Halo}
\label{subsec:ci-s}

\begin{figure}
\epsscale{1.1}
\plotone{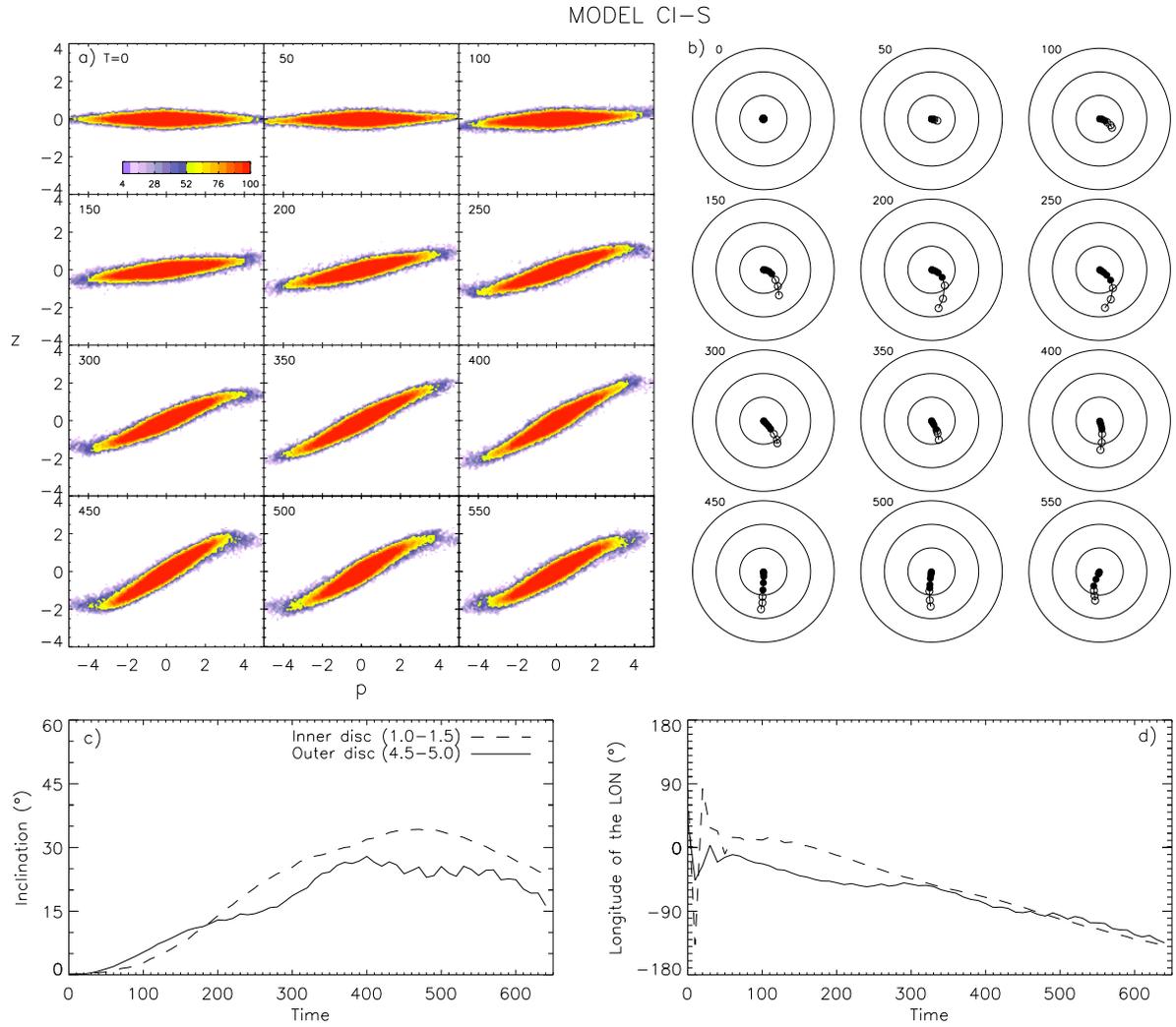}
\caption
{\label{fig:ci-s}Same as Figure \ref{fig:td-o}, but for model CI-S.  The
10 equal-size radial bins for panel b) range from $r=1.0$ to 6.0, each bin
with a width of 0.5, instead of 0.4.}
\end{figure}

Figure \ref{fig:ci-s} shows the evolution of the disk in a spherical
halo (model CI-S).  This model is to be compared with the axisymmetric and
triaxial halo cases.
As the halo in our model CI-S is perfectly spherical, the evolution of
the outer disk is driven by the interaction with the torus and with the inner
disk only.  The tilted torus gives an effect similar to an oblate halo to
the disk---it exerts a negative $T_L$ on the disk and makes the disk precess
in a retrograde sense about the axis normal to the torus plane.
But unlike in model TD-O, the precession rate caused by the torus
is an increasing function of $R$ (see the Appendix A of
Shen \& Sellwood 2006 and our Appendix B, but note that the precession
rate given in the former is along the torus plane while that in the latter
is along the equatorial plane), resulting in a trailing spiral of LON.
Thus the inner halo exerts a positive $T_P$ on the outer disk, and the
inclination of the outer disk decreases forming a Type II warp.

The inclination relative to the equatorial plane evolves significantly
with time (see panel $c$), but the inclination of the disk relative
to the torus plane stays constantly (this phenomenon is to be compared
with those in non-spherical halo models below).  The increase and decrease
of $i$ are well explained by equation (\ref{rate1t}).

\subsection{Oblate and Oblate-Like Halos}
\label{subsec:ci-o}

\begin{figure}
\epsscale{1.1}
\plotone{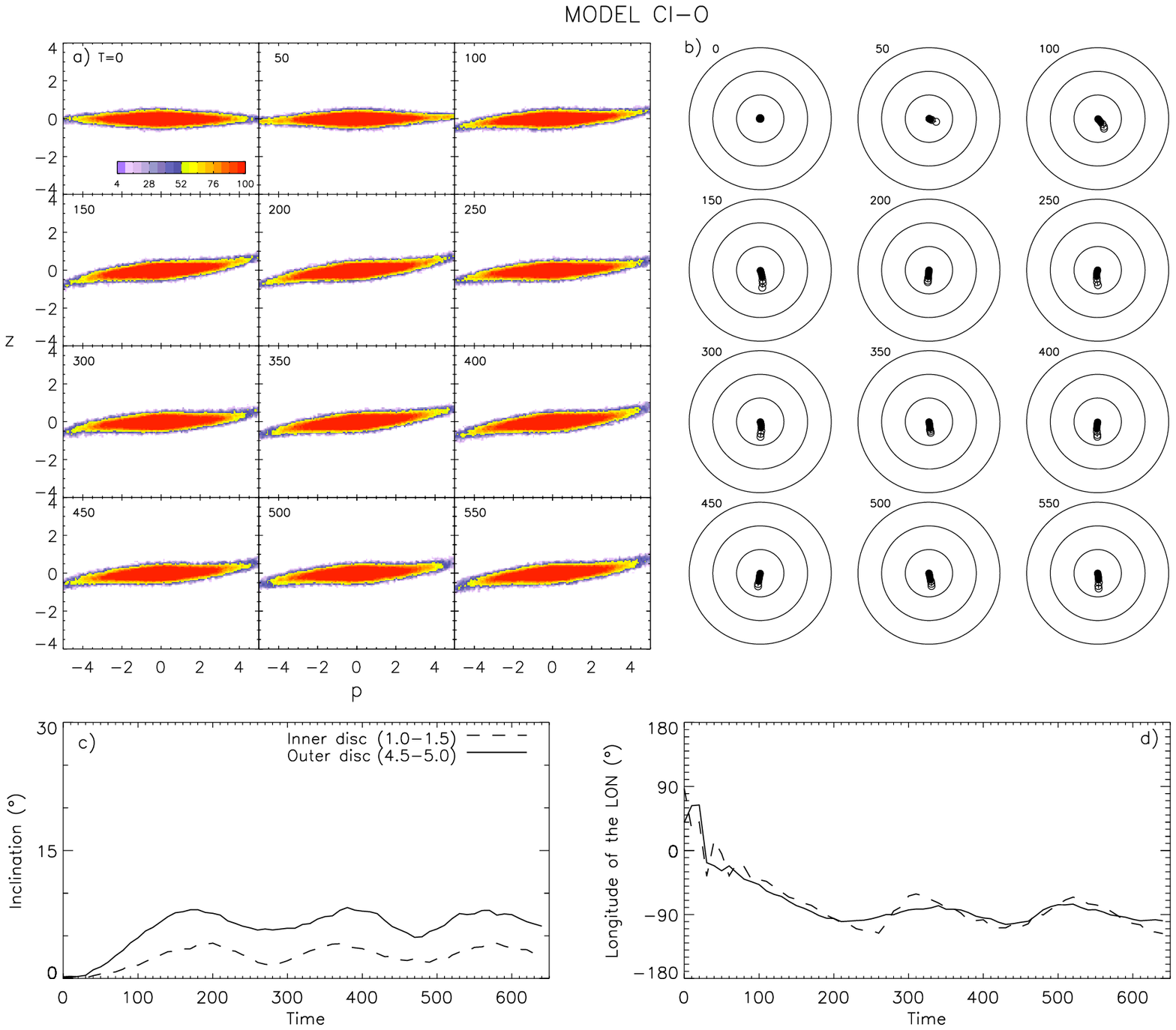}
\caption
{\label{fig:ci-o}Same as Figure \ref{fig:ci-s}, but for model CI-O.}
\end{figure}

\begin{figure}
\epsscale{0.5}
\plotone{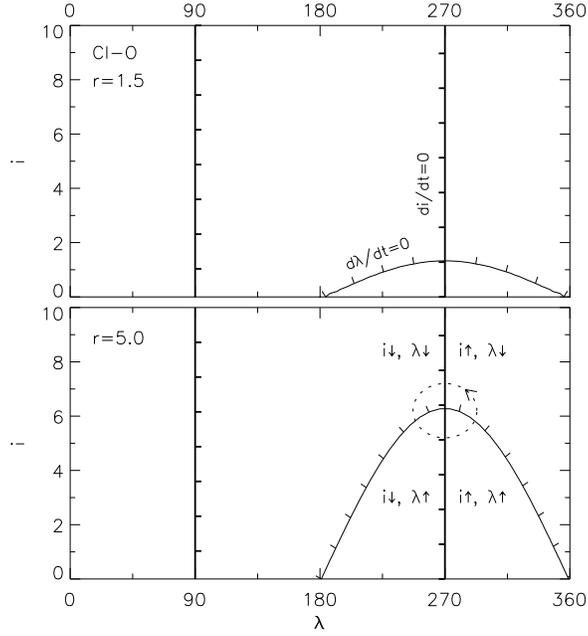}
\caption
{\label{fig:ci-o_rate} $di/dt=0$ (thick lines) and $d\lambda /dt=0$ (thin
lines) curves for
the inner disk ($r=1.5$; upper panel) and the outer disk ($r=5.0$;
lower panel) of model CI-O.  For $di/dt$ and $d\lambda /dt$, contributions
from both halo (eqs. [\ref{rate1}] \& [\ref{rate2}]) and torus
(eqs. [\ref{rate1t}] \& [\ref{rate2t}]) are considered.  Tickmarks
are drawn on the side of negative values (negative time derivatives).
The upward arrow indicates that the variable increases in that region,
while the downward arrow indicates that the variable decreases.  The
dashed circle with an arrow is centered at the ``equilibrium point''
(see the text for its definition) and illustrates how ($i$, $\lambda$) of
the disk would circulate around the equilibrium point.}
\end{figure}

Figure \ref{fig:ci-o} shows the evolution of the disk in an oblate halo
(model CI-O).  Since the oblate halo does not have a $J_{22}$ component,
the change of $i$ is caused only by the accreting torus.  Once the $i$
value departs from zero, $\lambda$ first moves toward $-90 \degr$ and
then both $i$ and $\lambda$ oscillate.  The oscillatory behaviors are
found to be drived by combined torques from the non-spherical halo and
the tilted torus.  Figure \ref{fig:ci-o_rate} shows four different
sign combinations of $di/dt$ and $d\lambda /dt$ from the combined
torques, and the arrangement of the sign combinations is in such a way
that the evolution of a point on the $i$--$\lambda$ plane follows a
round, counterclockwise revolution about the equilibrium point (the point
where both $di/dt$ and $d\lambda /dt$ are equal to zero).
Note that the center of $\lambda$ oscillation, $-90 \degr$, coincides
with the longitude of the ascending node of the torus, but the inclinations
of both outer and inner disks are somewhat smaller than that of the torus.

Inclinations of both inner and outer disks oscillate, but since the
equilibrium point is located at a higher $i$ for a larger $r$,
the mean $i$ value of the outer disk is higher than that of the inner disk,
and thus the disk forms a type-I warp, the opposite of the spherical halo
case.  Thus in oblate halos, the magnitude of the warp or $i_{rel}$
is determined by the radial gradient of the equilibrium inclination, and
is found to be smaller than that in the spherical halo model.
Unlike in model TD-O, the $\lambda$ values of the inner and outer disks in
model CI-O are not significantly separated.  This is because in model CI-O,
the inclinations of both inner and outer disks are so small that the $\cos i$
factors in equations (\ref{rate2}) and (\ref{rate2t}) are not so effective
in making the two $\lambda$'s separated as in model TD-O (i.e., the $\cos i$
values of the inner and outer disks are similar).

\begin{figure}
\epsscale{1.1}
\plotone{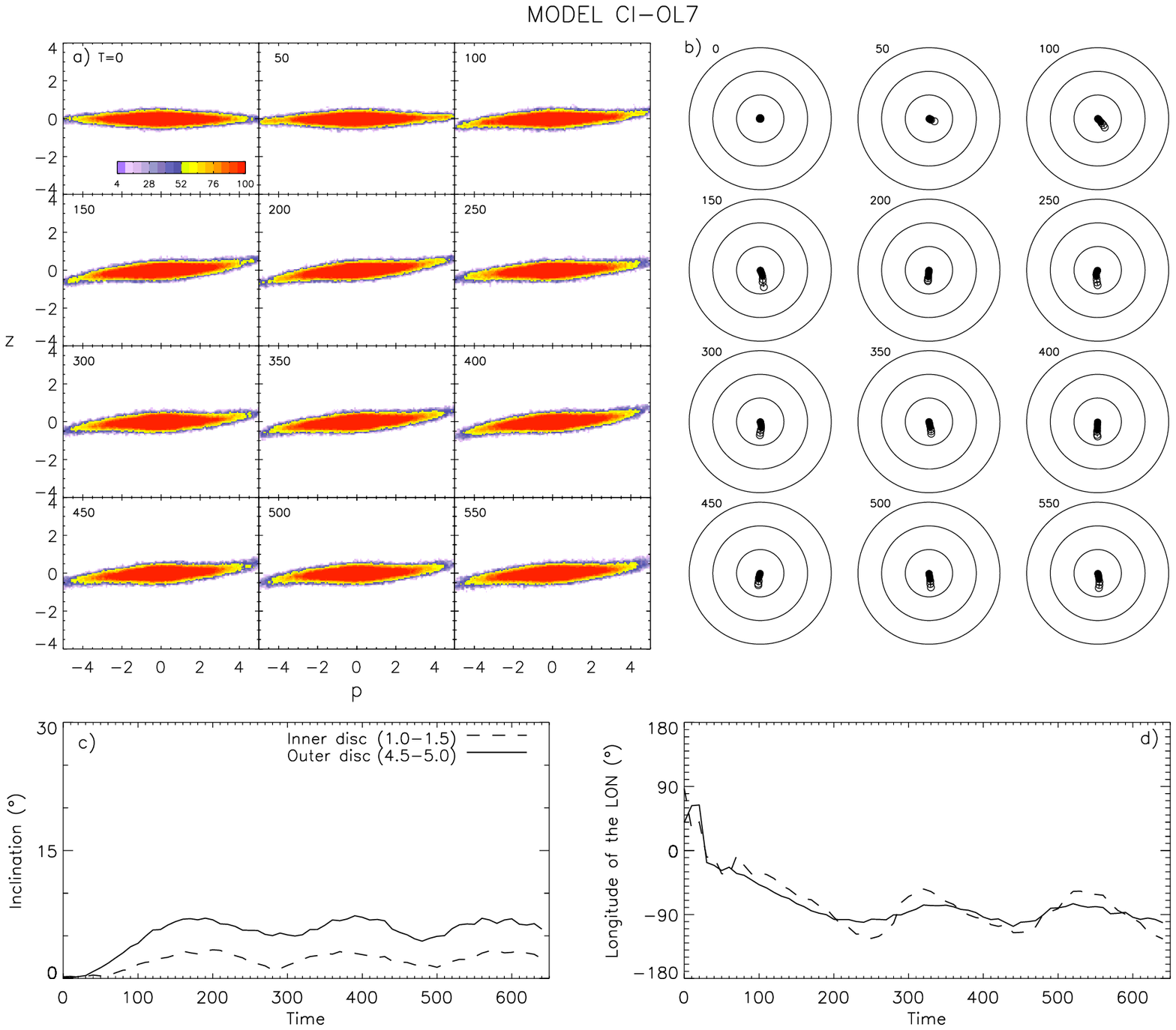}
\caption
{\label{fig:ci-ol7}Same as Figure \ref{fig:ci-s}, but for model CI-OL7.}
\end{figure}

\begin{figure}
\epsscale{0.5}
\plotone{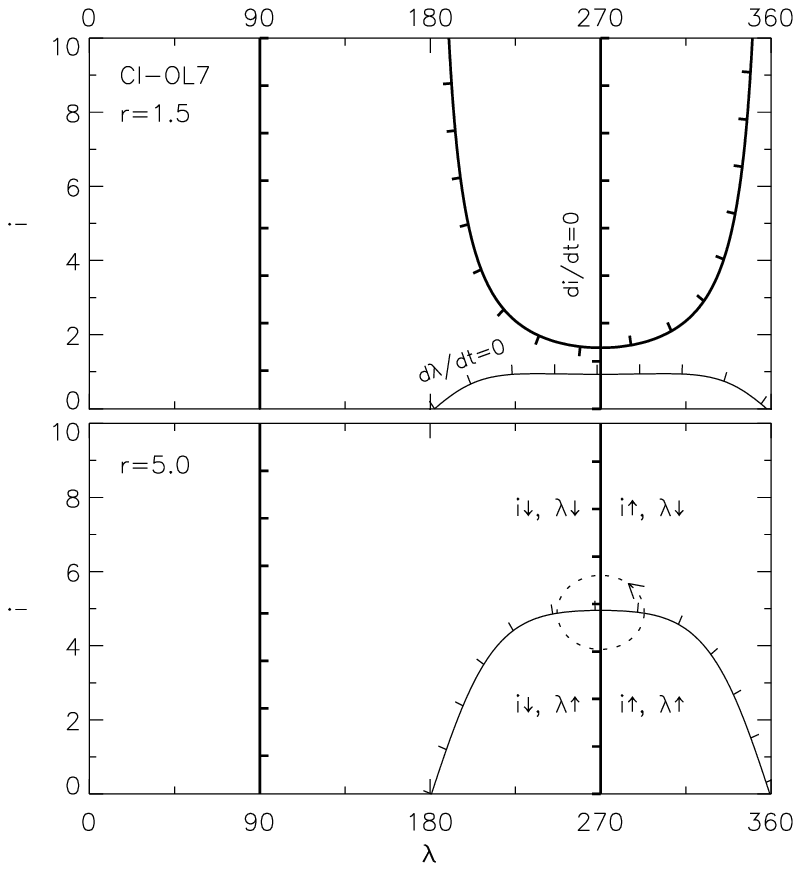}
\caption
{\label{fig:ci-ol7_rate}Same as Figure \ref{fig:ci-o_rate}, but for model
CI-OL7.}
\end{figure}

Figure \ref{fig:ci-ol7} shows the evolution of the disk in one of our
oblate-like halos (model CI-OL7).  In this model, the shorter of the two
halo axes in the equatorial plane aligns with the LON of the torus,
thus the oblate-like halo induces a positive $di/dt$ when $\lambda$ is in
the fourth quadrant ($270 \degr \la \lambda \la 360 \degr$) and a negative
$di/dt$ when $\lambda$ is in the third quadrant ($180 \degr \la \lambda
\la 270 \degr$).  Such $\lambda$ dependence of $di/dt$ is opposite to that
by the torus, but the contribution on $di/dt$ by the oblate-like halo is
much smaller than that by the torus because of the small value of $\sin i$
in equation (\ref{rate1}).  The average inclinations of both inner and
outer disks are smaller than those in the oblate halo model because the
equilibrium inclination is smaller for the oblate-like halo model ($J_{22}
\cos 2\lambda$ in eq. [\ref{rate2}] is negative when $\lambda$ is near
the equilibrium $\lambda$, $270 \degr$; see Fig. \ref{fig:ci-ol7_rate}).
Other than these lowered equilibrium inclinations for both inner and outer
disks, the overall evolution of the disk in model CI-OL7 is very similar
to that in model CI-O.

We find that the evolution of the disk in model CI-OL7b (figure not presented),
where the longer, not the shorter, of the two halo axes in the equatorial plane
aligns with the LON of the torus, is very similar to that of model CI-O as
well, implying that the orientation of the tilt of the torus relative to the
orientation of the oblate-like halo is not an important factor in determining
the configuration and the magnitude of the warp.  The only noticeable
difference between models CI-OL7b and CI-OL7 is the higher average $i$
values for both inner and outer disks of the former model because of the
positive $J_{22} \cos 2\lambda$ value in equation (\ref{rate2}) for
$\lambda \sim 270 \degr$.  Note that, however, all of our oblate and
oblate-like halo models have nearly the same magnitudes of the warp
(i.e., nearly the same $i_{rel}$).

\subsection{Prolate and Prolate-Like Halos}
\label{subsec:ci-p}

\begin{figure}
\epsscale{1.1}
\plotone{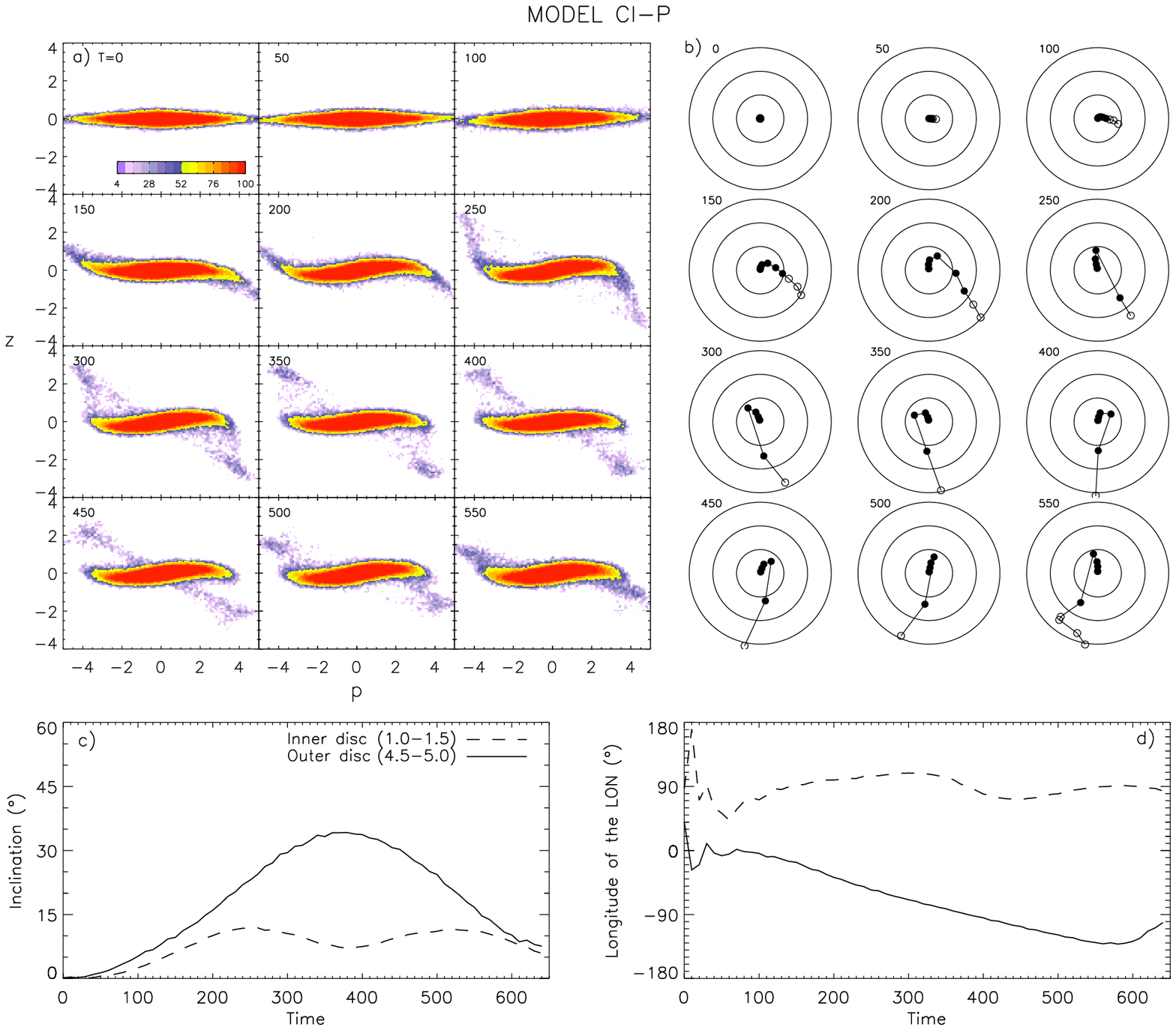}
\caption
{\label{fig:ci-p}Same as Figure \ref{fig:ci-s}, but for model CI-P.}
\end{figure}

Figure \ref{fig:ci-p} shows the evolution of the disk in a prolate
halo (model CI-P), which is rather different from those in spherical and
oblate halos. In a prolate halo, the magnitude of the warp is much larger,
and the outermost part of the disk is even detached from the rest of the
disk when the LONs of the outermost part and the main disk are apart by
$\sim 180 \degr$.

\begin{figure}
\epsscale{0.5}
\plotone{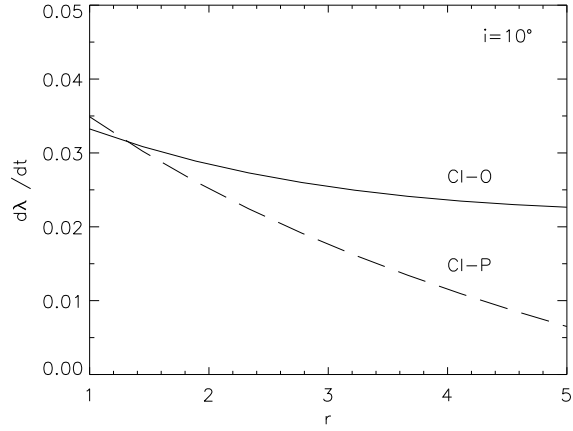}
\caption
{\label{fig:ci-p_rate_r}Radial profiles of $d\lambda /dt$ for models CI-O
(solid line) and CI-P (dashed line) when $i=10 \degr$.  Contributions
from both halo (eq. [\ref{rate2}]) and torus (eq. [\ref{rate2t}]) are
considered.  $d\lambda /dt$ of model CI-P varies more significantly
than that of model CI-O between $r=1$ and 5, which is thought to be the main
cause of the independent evolution of the outer disk from the inner disk.}
\end{figure}

\begin{figure}
\epsscale{0.5}
\plotone{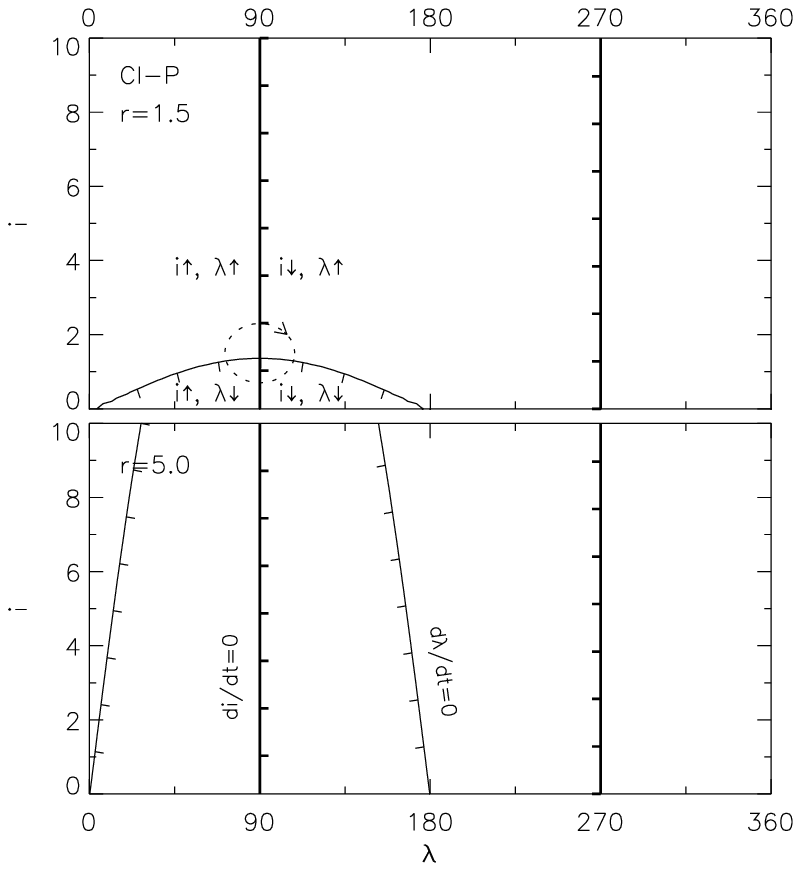}
\caption
{\label{fig:ci-p_rate}Same as Figure \ref{fig:ci-o_rate}, but for model CI-P.}
\end{figure}

This rather large difference is thought to be because of the strong radial
dependence of the combined $d\lambda /dt$ from the halo and the torus.
Figure \ref{fig:ci-p_rate_r}
shows that the combined $d\lambda /dt$ for the prolate halo has a
significantly stronger radial dependence than that for the oblate halo.
Thus unlike in the oblate halo, the outer disk in the prolate halo can
evolve separately from the inner disk from the beginning of the simulation.
The LON of the inner disk oscillates around the equilibrium
$\lambda$, $90 \degr$ (see Fig.  \ref{fig:ci-p_rate}), whereas the LON of
the outer disk, which is more affected by the self-gravity within the disk
than the inner disk is, appears to experience more complicated evolution.
In the beginning of the simulation ($T \lsim 150$), the combined $d\lambda /dt$
on the outer disk from the halo, torus, and the inner disk is nearly zero,
so $\lambda$ of the outer disk barely evolves.  But once $\lambda$'s of
the inner and outer disks are apart by more than $90 \degr$, the second
term in equation (\ref{rate2d}) becomes negative.  This then breaks the
subtle equilibirum and makes the outer disk precess in a retrograde fashion.

Unlike in model TD-P, the $\lambda$ values of the inner and outer disks in
model CI-P are well separated.  In model TD-P, the $\cos i$ factor in
$d\lambda /dt$ (eq. [\ref{rate2}]) causes the outer disk, which has a smaller
$i$ than the inner disk, to catch up the $\lambda$ of the inner disk.
But in model CI-P, the inclinations of the inner and outer disks are both
so small in the beginning of the simulation that the $\cos i$ factor
in the combined $d\lambda /dt$ from the halo and the torus is not so
effective in making the two $\lambda$'s move together ($|d\cos i/di|$ is
smaller for small $i$'s).

Because $\lambda$ of the inner disk is initially $\sim 90 \degr$ ahead
(in the sense of disk spin) of that of the outer disk, the inner disk
experiences a positive
$di/dt$ from the outer disk (see eq. [\ref{rate1d}]), thus has an equilibrium
$i$ value that is higher than the one determined by considering the
torques from the halo and torus only as in Figure \ref{fig:ci-p_rate}.
The $i$ value of the outer disk increases in the first half of the simulation
and decreases in the second half of the simulation because the combined
$di/dt$ by the halo and the torus is positive when $\lambda$ is in the
fourth quadrant and negative when $\lambda$ is in the third quadrant (see Fig.
\ref{fig:ci-p_rate}).

While the evolution of both inner and outer disks in spherical and oblate
halos can be explained by the combined torques from the halo and the torus
only, in a prolate halo, the self-gravity within the disk plays
a non-negligible role in the $i$ and $\lambda$ evolutions of the disk.
This is because in a prolate halo, 1) the $di/dt$ term from the inner disk
is not negligible as the LONs of the inner and outer disks are well separated
from each other (see eq. [\ref{rate1d}]), and 2) the $d\lambda /dt$
term from the inner disk is relatively more important as the $J_{20}$ values
of the halo and the torus have opposite signs.

\begin{figure}
\epsscale{1.1}
\plotone{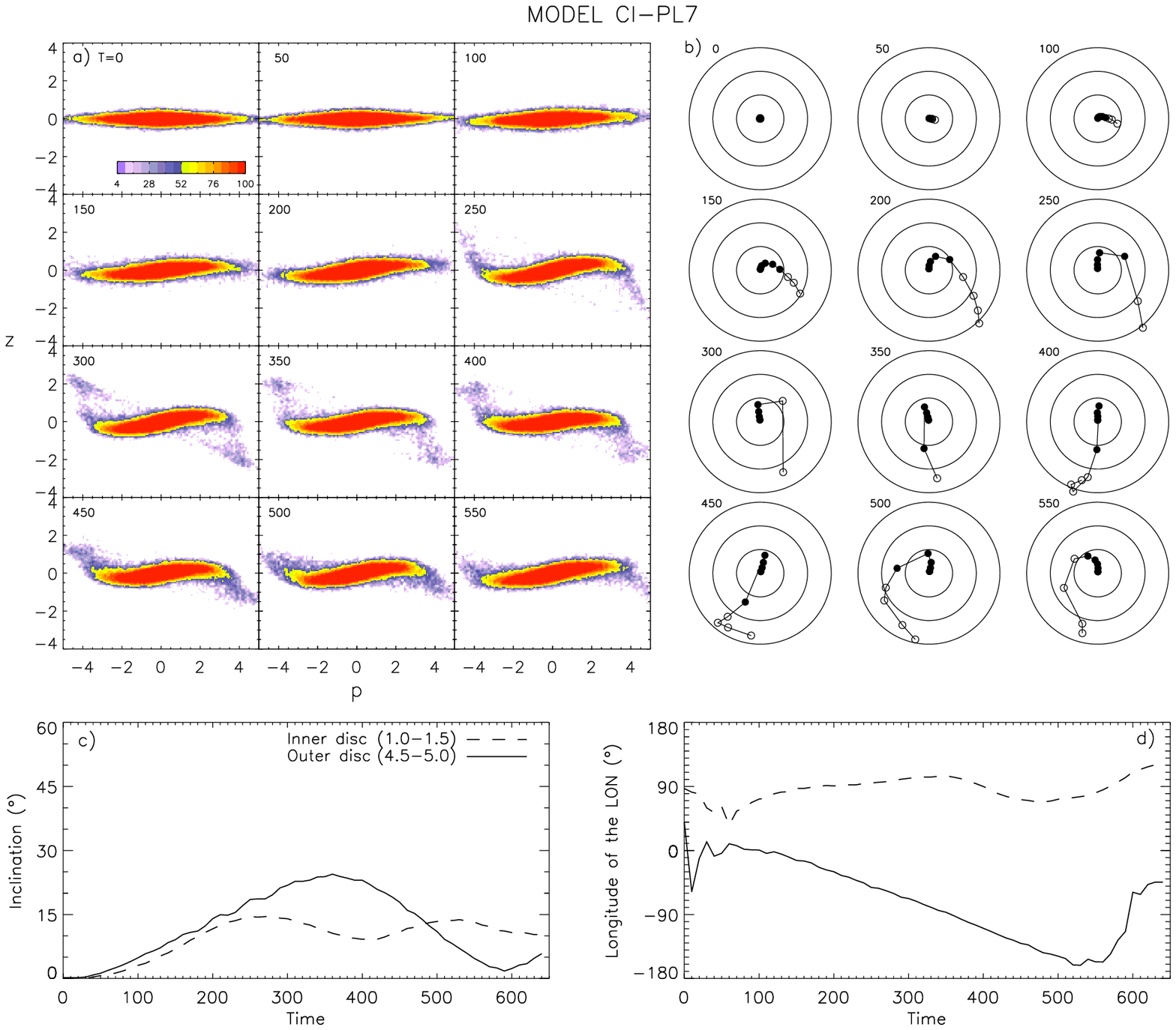}
\caption
{\label{fig:ci-pl7}Same as Figure \ref{fig:ci-s}, but for model CI-PL7.}
\end{figure}

Figures \ref{fig:ci-pl7} shows the evolution of the disk in one of our
prolate-like halo models (model CI-PL7), where the shorter of the two
halo axes in the equatorial plane aligns with the ascending LON of the
torus.  The evolution of CI-PL7 is generally similar to that of CI-P
except that the magnitudes of the warp and the detachment of the outer disk
less significant than those in model CI-P.  This is because the $J_{22}$
component of this triaxial halo induces a negative $di/dt$ on to the outer
disk when $\lambda$ is in the fourth quadrant (i.e., during the first half
of the simulation), causing the peak $i$ that the outer disk acquires to
be smaller than that in model CI-P.

\begin{figure}
\epsscale{1.1}
\plotone{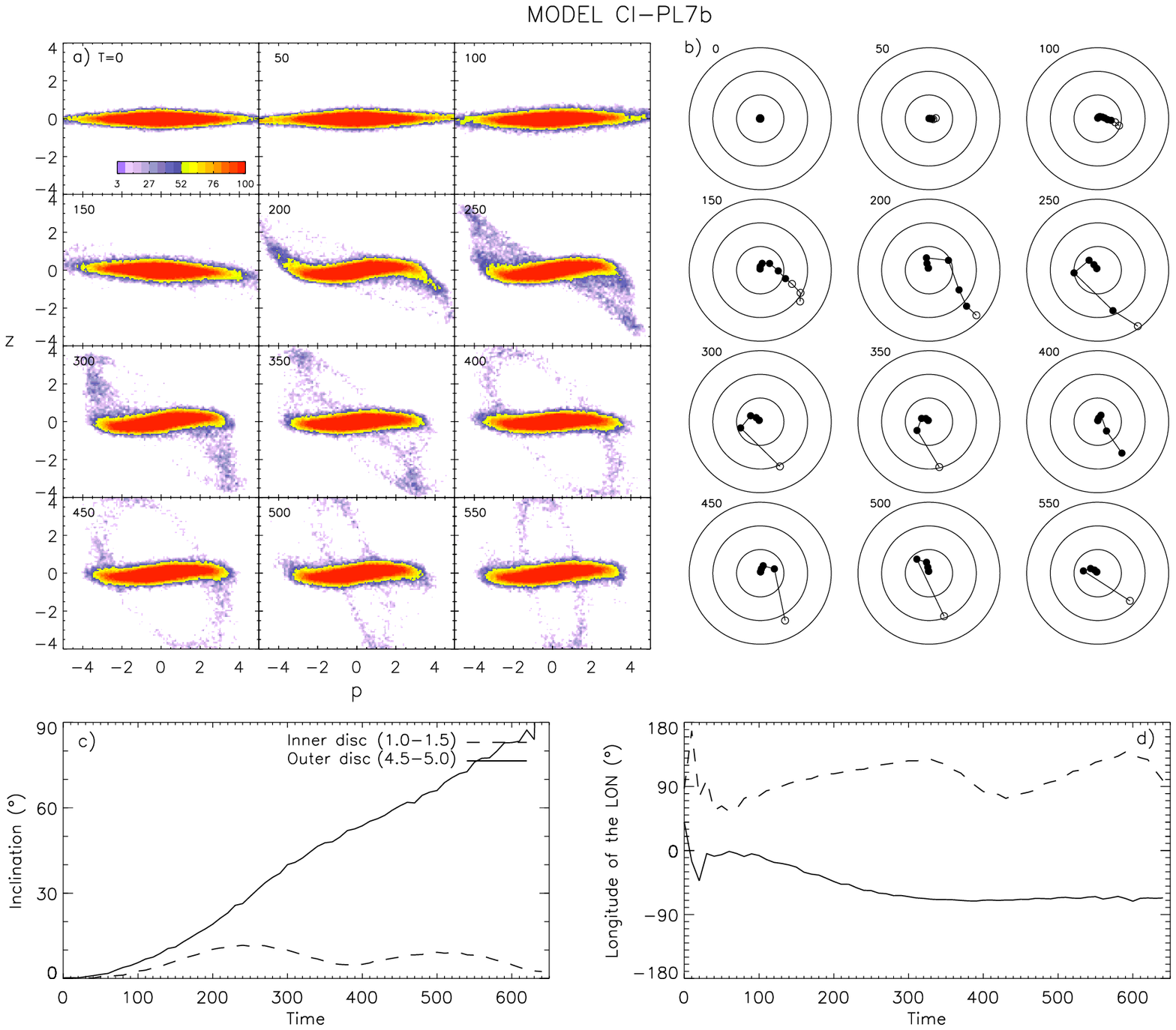}
\caption
{\label{fig:ci-pl7b}Same as Figure \ref{fig:ci-s}, but for model CI-PL7b.}
\end{figure}

\begin{figure}
\epsscale{0.5}
\plotone{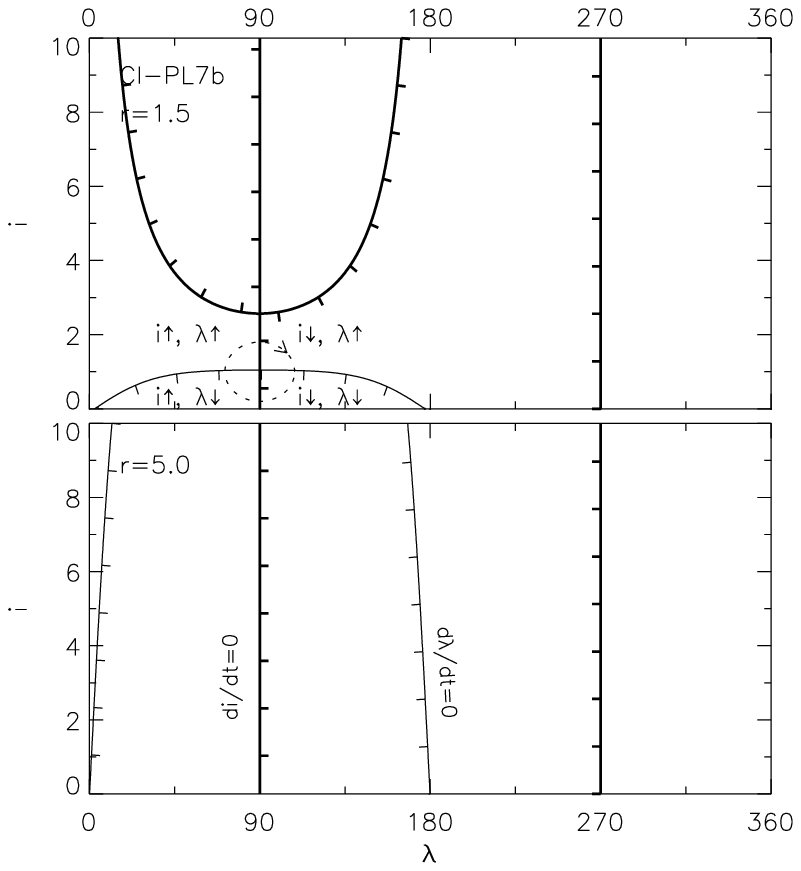}
\caption
{\label{fig:ci-pl7b_rate}Same as Figure \ref{fig:ci-o_rate}, but for model
CI-PL7b.}
\end{figure}

Figures \ref{fig:ci-pl7b} shows the evolution of the disk in another
prolate-like halo (models CI-PL7b), where the longer of the two
halo axes in the equatorial plane aligns with the ascending LON of the
torus.  The second term in equation (\ref{rate2}) now causes a positive
(instead of negative as in model CI-PL7) $d\lambda/dt$ for the outer disk
when $\lambda \sim 270 \degr$, and this apparently balances the negative
$d\lambda/dt$ from the inner disk when the $\lambda$'s of the inner and outer
disks are apart by more than $90 \degr$ so that $\lambda$ of the outer disk
stays near $290\degr$ after $T \simeq 200$.  At this equilibrium $\lambda$,
the combined $di/dt$ on the outer disk from the halo and the torus is positive
(see Fig. \ref{fig:ci-pl7b_rate}).  Consequently, $i$ keeps increasing even
after the outer disk is completely detached from the inner disk, and
eventually the system resembles polar-ring galaxies.

While the evolutions of the disks in oblate and oblate-like halos are
alike and can be described rather simply and coherently without a
consideration of the self-gravity within the disk, those in prolate
and prolate-like halos sensitively depend on the relative importances
between the torques from the halo, the torus, and the disk itself, as well
as on the relative orientation between the halo and the torus.  The
two-disk analysis that we adopted to describe the effect of the self-gravity
is found to give qualitatively correct descriptions, but does not give
quantitatively accurate estimates for the torques.  Thus it is rather
difficult to generally describe the evolution of the disk when the
self-gravity is not negligible, which is the case of the disks in
prolate and prolate-like halos.

\section{SUMMARY AND DISCUSSION}
\label{sec:summary}

We have performed a series of simulations of galactic warps for two
scenarios: 1) stellar disks that are initially tilted relative to the
equatorial plane of triaxial halos, and 2) stellar disks in axisymmetric
or triaxial halos with late infall of dark matter that is modeled as
a torus tilted relative to the equatorial plane of the disk and the halo.

In our initially tilted disk (TD) simulations, the triaxiality in the oblate
halo does not significantly alter the characteristics and magnitude of the
warp, although the latter oscillates as the disk precesses and is smaller for
a larger triaxiality.  Moderate ($7~\%$) triaxiality of the prolate halo
results in the warp similar to that in the axisymmetric prolate halo with
an oscillatory behavior, while a larger ($13~\%$) triaxiality makes the disk
precess about the minor axis of the halo, instead of the major axis that the
disk initially precessed about, and considerably reduces the magnitude of
the warp.

In our cosmic infall (CI) simulations, the magnitude of the warp in the
non-spherical halo is determined by the difference in the ``equilibrium $i$''
between the inner and outer disks.  While the magnitude of the warp
in oblate and oblate-like halos is smaller than that in the spherical halo,
in prolate and prolate-like halos the warp can grow larger and the outermost
part of the disk can even detach from the main disk body.  For a prolate-like
halo system where the ascending LON of the dark matter torus aligns with
the longer of the two halo axes in the equatorial plane, the detachment
becomes more extreme and the system resembles polar-ring galaxies.
Note that, however, cosmological simulations show that the angular momentum
vector of the halo preferentially aligns with the minor axis of the halo
(see the references below), thus the disk morphologies found in our prolate
and prolate-like halo simulations (detached outer disk and polar-ring
configuration) are not expected to take place often.

Regardless of the warp scenario, the triaxiality of the halo causes
$i$ and $\lambda$ to oscillate (i.e., it causes the disk to nutate).
the differential nutation between the inner and outer disks generally
makes the magnitude of the warp slightly diminish and fluctuate, but
this could also cause the outermost part of the disk to completely detach
from the main body if this effect is boosted by the torque from the dark
matter torus as in model CI-PL7b.

In the Appendices we derived the approximate formulae for the torques exerted
on the disk by the triaxial halo and the dark matter torus, and with
these formulae, we have successfully described the $i$ and $\lambda$
evolutions of the disks.  The characteristics of the halo and
the torus are represented by three parameters $J_{20}$, $J_{21}$, and
$J_{22}$, which can be useful in
predicting the evolution of a disk in various different halo and torus
models, or in estimating the relative importance between the halo and
the torus.

The techniques used in deriving these formulae could be applied for halos
with more complex structures.  Bailin \& Steinmetz (2005) found from their
$\Lambda$CDM $N$-body simulations that the principal axes of the galaxy
dark matter halos are internally misaligned (i.e., the directions of the
principal axes are functions of galactocentric radius).  In such systems,
although the stellar disk will quickly be in equilibrium with the inner halo,
it may experience torques from the misaligned outer halo for an extended
period, and the higher order terms than $J_{21}$ and $J_{22}$ might be
important. Such configuration could be responsible for U- or L-shaped
warps, and for these cases, each radial bin of the disk could be modeled
with two broken half-rings, instead of a whole ring.

The main goal of the present study is to understand the fundamental
dynamics between the disk and the non-spherical halo (with and without
the dark matter torus), thus we have considered models that result
in relatively prominent warps (a large initial tilt angle of the disk
and a large non-sphericity of the halo).  However, most of the observed
edge-on disk galaxies have only a hint of warped structure, if any, near
$R_d \simeq 4$--5 in their stellar disks.
For example, Ann \& Park (2006) found that out of their
325 sample galaxies 165 (51~\%) have an S-shaped warp, and 142 (86~\%) of
these S-shaped warps have warp angles $\alpha$ smaller than $5 \degr$
($\alpha$ is the inclination angle of the tip of the outermost isophotes
from the mean disk major axis, and is very close to our $i_{rel}$).

\begin{figure}
\epsscale{0.75}
\plotone{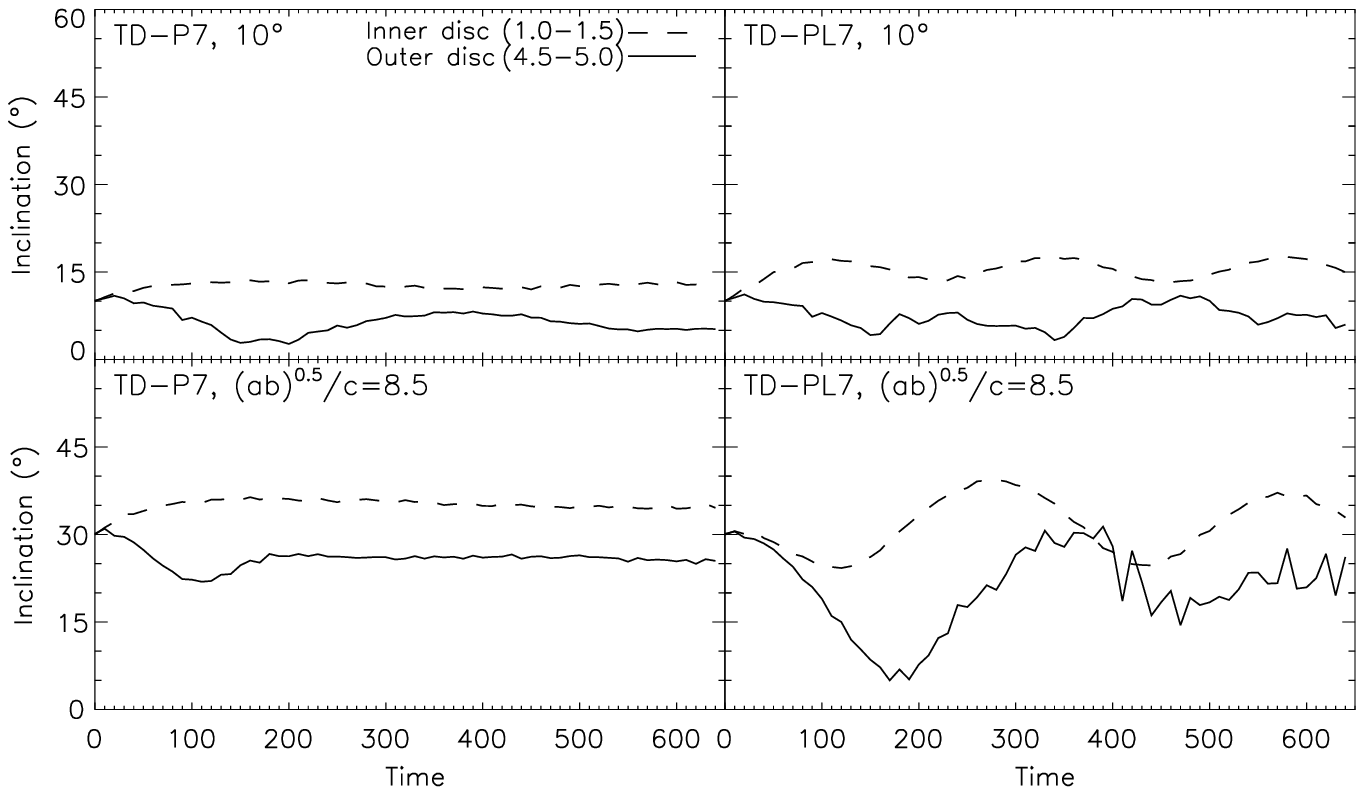}
\caption
{\label{fig:td-other}$i$ evolution of the two radial bins for models
that are variants from models TD-P7 (left panels) and TD-PL7 (right panels).
The upper panels are the models with an initial disk inclination of $10 \degr$
instead of $30 \degr$, and the lower panels are the models with an bulge
axis ratio $\sqrt{ab}/c$ of 8.5 instead of 7.5.  These variant models result
in smaller magnitudes of warping than our canonical models, TD-P7 and TD-PL7.}
\end{figure}

Our simulations for the initially tilted disk scenario show that prolate
and prolate-like halo models can excite more prominent warps than
oblate and oblate-like halo models.  However, this does not necessarily
imply that, as Ideta et al. (2000) interpreted, oblate and oblate-like
models are not appropriate to explain the observed warps--they may not
be able to sustain a prominent warp, but surely can resemble the majority
of the observed warps, i.e., those with small $\alpha$ values.  On the other
hand, prolate and prolate-like models can explain warps with both
large and small $\alpha$ values by having different initial tilt angles
or halo axis ratios.  Figure \ref{fig:td-other} shows that $i_{rel}$ can
become smaller if prolate and prolate-like models have an initial tilt angle
of $10 \degr$, instead of $30 \degr$, or a halo axis ratio of $\sqrt{ab}/c=
0.85$, instead of 0.75.


Our simulations for the tilted cosmic infall scenario also show that prolate
and prolate-like halo models can excite more prominent warps than oblate
and oblate-like halo models.  The outermost disk in those models can
detach from the main disk body or even form a polar ring at least for some
time, but such systems are barely observed.  The magnitude of the warp
is suppressed in oblate and oblate-like halos, and is even smaller than
in the spherical halo.  Thus for the cosmic infall scenario, the way to
obtain a nice and clean, prominent warp would be to have a (nearly) spherical
halo and/or dark matter infall that gives a large $J_{21}$ value.

These results appear to be consistent with the recent $\Lambda$ cold dark
matter cosmological simulations.  Bailin et al. (2005) found in their
hydrodynamic simulations that the (inner) halo minor axes are nearly aligned
with the disk axis, and Bett et al. (2007) analyzed the Millennium simulation
(Springel et al. 2005) to find that the majority of halos have their
angular momentum vector aligned with their minor axis.  Both of these,
which confirm the findings of earlier studies by Dubinski (1992) and
Warren et al. (1992), imply that most of the disks are situated in an
oblate or oblate-like halo,
and can explain why 1) most of the observed warps are quite weak (warps are
suppressed in oblate and oblate-like halos) and 2) disks with detached
outer part or polar-ring systems are not observed or rare.

\acknowledgements
We are grateful to Juhan Kim, Woong-Tae Kim, Jounghun Lee, Juntai Shen, and
Linda Sparke for helpful discussion.  We thank the anonymous referee for
valuable comments.  M. J. was supported by the Graduate School of Kyung
Hee University through 2007 Excellent Research Paper Scholarship.
S. S. K. and H. B. A. were supported by the Astrophysical Research Center for
the Structure and Evolution of the Cosmos (ARCSEC) of the Korea Science and
Engineering Foundation through the Science Research Center (SRC) program.
Simulations presented in this paper were performed on the GRAPE cluster at
Kyung Hee University and on the Linux cluster at Korea Astronomy \& Space
Science Institute (KASI), which was built with the fund from ARCSEC and KASI.

\appendix

\section{Approximate Torques by the Triaxial Halo}
\label{appen:halo}

Here we derive approximate torques exerted on a ring of disk material
by triaxial halos following the procedure in Steiman-Cameron \& Durisen
(1984), who considered the halo mass distribution inside the ring only.
We extend their work and consider the halo mass distribution
both inside and outside the radius of the ring.  We then derive the rates
of changes in inclination and LON of the ring due to the non-spherical
components of the halo.

Gravitational potential from a mass distribution $\rho$ can be written
as an expansion in spherical harmonics $Y_{lm}$,
\begin{equation}
	\Phi(r,\theta,\phi) = -4 \pi G \displaystyle\sum_{l=0}^\infty
                              \displaystyle\sum_{m=-l}^l
                              \frac{Y_{lm}(\theta,\phi)}
                              {2 l+1} \biggl[\frac{1}{r^{l+1}}
                              \int_{0}^{r} \rho_{lm} (r') r'^{l+2}dr'
                              + r^l \int_{r}^{\infty} \frac{\rho_{lm}
                              (r')}{r'^{l-1}}dr'\biggr].
\end{equation}
Here $\rho_{lm}$ are the $\rho$-coefficients of the shell lying between
$r$ and $r+\delta r$,
\begin{equation}
	\rho_{lm}(r')= \int Y_{lm}^\ast (\theta,\phi) \rho (r',\theta,\phi)
                       d\Omega,
\end{equation}
where $Y_{lm}^\ast$ are the spherical Legendre conjugate functions, and
$d\Omega$ is the solid angle element.  Then the first-order corrections
to the potential can be written as
\begin{equation}
\label{Phi1}
	\Phi_1 = -\frac{G}{4} \left[ J_{20} (3\cos^2 \theta - 1)
                 + 3J_{22} \sin^2 \theta \cos 2\phi \right ],
\end{equation}
where $J_{20}$ and $J_{22}$ are defined by
\begin{eqnarray}
\label{coef}
	J_{20}(r) & = & J^{int}_{20}(r)+J^{ext}_{20}(r) \nonumber \\
	J_{22}(r) & = & J^{int}_{22}(r)+J^{ext}_{22}(r)
\end{eqnarray}
and
\begin{mathletters}
\label{coef20}
\begin{eqnarray}
	J^{int}_{20}(r) & = & \frac{1}{r^3} \int_{r'=0}^{r'=r}
					     (2z'^2-x'^2-y'^2) \rho dV'      \\
	J^{ext}_{20}(r) & = & r^2 \int_{r'=r}^{r'=\infty} (2z'^2-x'^2-y'^2)
                              \frac{\rho}{r'^5} dV'
\end{eqnarray}
\end{mathletters}
\begin{mathletters}
\label{coef22}
\begin{eqnarray}
	J^{int}_{22}(r) & = & \frac{1}{r^3}
                              \int_{r'=0}^{r'=r} (x'^2-y'^2) \rho dV'        \\
	J^{ext}_{22}(r) & = & r^2 \int_{r'=r}^{r'=\infty} (x'^2-y'^2)
                              \frac{\rho}{r'^5} dV'.
\end{eqnarray}
\end{mathletters}
Oblate halos have negative $J_{20}$ values while prolate halos have positive
$J_{20}$'s.  Also note that oblate and prolate halos have $J_{22}=0$, and
triaxial halos with $a>b$ ($b>a$) have positive (negative) $J_{22}$'s.

Now with equations (A16) to (A18) of Steiman-Cameron \& Durisen (1984),
the average first-order potential corrections for a ring of disk material
that has a radius of $r$, inclination $i$, and the longitude of ascending
LON $\lambda$ become
\begin{equation}
\label{Phi1_r}
	\langle \Phi_1 \rangle = -\frac{G}{8} \left[ J_{20} (3\sin^2 i - 2)
                                 +3J_{22} \cos 2\lambda \sin^2 i\right].
\end{equation}
If the spin of the ring is much faster than the precession or the inclination
change of the ring, the Lagrange equations of motion for the ring become
\begin{eqnarray}
\label{lagrange1}
	\frac{di}{dt}       & = & - \frac{1}{r V_c \sin i}
                                    \frac{\partial \langle \Phi_1 \rangle}
                                         {\partial \lambda}                  \\
\label{lagrange2}
	\frac{d\lambda}{dt} & = &   \frac{1}{r V_c \sin i}
                                    \frac{\partial \langle \Phi_1 \rangle}
                                         {\partial i}
\end{eqnarray}
(Habe \&  Ikeuchi 1985), where $V_c$, the circular (spin) velocity of the
ring, is defined to be positive when the ring spins counterclockwise when
viewed from the positive $z$-axis, and the Hamiltonian equations of motion
for the ring yield
\begin{eqnarray}
\label{hamiltonian}
	\dot p_i       & = & -\frac{\partial \langle \Phi_1(r,i,\lambda)
                                    \rangle}{\partial i}                     \\
	\dot p_\lambda & = & -\frac{\partial \langle \Phi_1(r,i,\lambda)
                                    \rangle}{\partial \lambda}
\end{eqnarray}
(Arnaboldi \& Sparke 1994).  Since the $L$-axis coincides with the LON,
we have
\begin{equation}
	T_L = \dot p_i
\end{equation}
and since the $P$-component angular momentum of the ring is $-mr V_c \sin i$,
we have
\begin{equation}
\label{tp_idot}
	T_P = -mr V_c \cos i \frac{di}{dt}
\end{equation}
(see \S \ref{sec:theory} for the definitions of $L$- and $P$- axes).
Now the $L$ and $P$ components of the torque become
\begin{eqnarray}
\label{torque1}
	T_L & = & -\frac{\partial \langle \Phi_1(r,i,\lambda) \rangle}
                        {\partial i}                               \nonumber \\
            & = & \frac{3G}{4} (J_{20}+J_{22}\cos 2\lambda) \sin i \cos i \\
\label{torque2}
	T_P & = & -\cot i \frac{\partial \langle \Phi_1(r,i,\lambda) \rangle}
                               {\partial \lambda}                  \nonumber \\
            & = & -\frac{3G}{8} J_{22}\sin 2\lambda \sin 2i.
\end{eqnarray}
Finally, from equations (\ref{lagrange1}), (\ref{lagrange2}), (\ref{torque1}),
and (\ref{torque2}), the rate of inclination change and the precession rate
can be given by
\begin{eqnarray}
\label{rate1}
        \frac{di}{dt}       & = & -\frac{T_P}{r V_c \cos i}  \nonumber \\
                            & = & \frac{3G}{4r V_c}
                                  J_{22}\sin 2\lambda \sin i                 \\
\label{rate2}
	\frac{d\lambda}{dt} & = & \frac{T_L}{r V_c \sin i}   \nonumber \\
	                    & = & \frac{3G}{4r V_c}
                                  (J_{20}+J_{22}\cos 2\lambda) \cos i.
\end{eqnarray}

\section{Approximate Torques by the Dark Matter Torus}
\label{appen:torus}

Here we derive approximate torques exerted on a ring of disk material
by the accreting torus of a mass $M_t$ that is inclined at
$i_t$ from the equatorial ($x$-$y$) plane of the halo with the ascending
LON aligned with the negative $y$-axis.  We assume that the torus has
an infinitesimally small cross section and a radius of $R_t$ that is
larger than that of the ring.

The potential by the torus does not have $l=1$ terms as in the case of
triaxial halos, but it does have $l=2, \, m=\pm 1$ terms besides
$l=2, \, m=0,\pm 2$ terms.  Then the first-order corrections to the potential
can be written as
\begin{equation}
\label{Phi1t}
	\Phi_1 = -\frac{G}{4} \left[ J_{20} (3\cos^2 \theta - 1)
                 + 6J_{21} \sin 2\theta \cos \phi
                 + 3J_{22} \sin^2 \theta \cos 2\phi \right ].
\end{equation}
We find that the coefficients $J_{20}$, $J_{21}$, and $J_{22}$ have the
following forms for the torus:
\begin{eqnarray}
\label{coef20t}
	J_{20}(r) & = & \frac{M_t}{2} \frac{r^2}{R_t^3} (3 \sin^2 i_t - 2) \\
\label{coef21t}
	J_{21}(r) & = & \frac{M_t}{4} \frac{r^2}{R_t^3} \sin 2 i_t         \\
\label{coef22t}
	J_{22}(r) & = & - \frac{M_t}{2} \frac{r^2}{R_t^3} \sin^2 i_t.
\end{eqnarray}

Now with equations (A16) to (A18) of Steiman-Cameron \& Durisen (1984),
the average first-order potential corrections for a ring of disk material
that has a radius of $r$, inclination $i$, and the longitude of ascending
LON $\lambda$ become
\begin{equation}
\label{Phi1_rt}
	\langle \Phi_1 \rangle = -\frac{G}{8} \left[ J_{20} (3\sin^2 i - 2)
                                 - 6J_{21} \sin \lambda \sin 2i
                                 + 3J_{22} \cos 2\lambda \sin^2 i \right ].
\end{equation}
Then from equations (\ref{torque1}) and (\ref{torque2}), the $L$ and $P$
component torques are given by
\begin{eqnarray}
        T_L & = & \frac{3G}{8} (J_{20} \sin 2i - 4J_{21} \sin \lambda \cos 2i
                  + J_{22}\cos 2\lambda \sin 2i ) \\
        T_P & = & \frac{3G}{8} (4 J_{21} \cos \lambda \cos^2 i
                  + J_{22} \sin 2\lambda \sin 2i ),
\end{eqnarray}
and as in Appendix \ref{appen:halo}, the rate of inclination change and the
precession rate are given by
\begin{eqnarray}
\label{rate1t}
        \frac{di}{dt}       & = & \frac{3G}{4r V_c} (
                                  2 J_{21} \cos \lambda \cos i
                                  + J_{22}\sin 2\lambda \sin i )   \nonumber \\
                            & \simeq & \frac{3G}{4r V_c}
                                  2 J_{21} \cos \lambda \cos i               \\
\label{rate2t}
	\frac{d\lambda}{dt} & = & \frac{3G}{4r V_c} \left (
                                  J_{20} \cos i
				  - 2 J_{21} \sin \lambda \frac{\cos 2i}{\sin i}
				  + J_{22} \cos 2\lambda \cos i \right )
                                                                   \nonumber \\
	                    & \simeq & \frac{3G}{4r V_c} \left (
                                  J_{20} \cos i
				  - 2 J_{21} \sin \lambda \frac{\cos 2i}{\sin i}
				  \right ).
\end{eqnarray}
Note that for our torus, $J_{22}$ terms are negligible compared to the other
terms (see Table \ref{table:jvalues}).

Shen \& Sellwood (2006) derives the precession rate of a disk due to an
accreting torus by assuming that the torus lies in the equatorial plane
and the disk is inclined at a certain angle relative to the equatorial
plane (note that our $i$ and $\lambda$ are defined to be the inclination
and the LON respectively in the coordinates where the torus is inclined
relative to the equatorial plane; when the torus lies
in the equatorial plane, the inclination of the disk does not change and
only $\lambda$ of the disk evolves).  The rates of changes in $i$ and
$\lambda$ can be also derived from the precession rate by Shen \& Sellwood,
\begin{equation}
\label{rate2ss}
	\frac{d\lambda'}{dt} = -\frac{3G M_t r}{4 R_t^3 V_c} \cos i',
\end{equation}
where $\lambda'$ is the LON of the disk along the torus plane and $i'$ is
the inclination of the disk relative to the torus plane.
By transforming the coordinates from the torus plane to the equatorial plane,
one obtains the following rates of changes:
\begin{eqnarray}
\label{rate1tt}
        \frac{di}{dt}       & = & -\cos \lambda \sin i_t \frac{d\lambda'}{dt} \\
\label{rate2tt}
	\frac{d\lambda}{dt} & = & \left ( \cos i_t + \cot i \sin \lambda
                                      \sin i_t \right ) \frac{d\lambda'}{dt}.
\end{eqnarray}
The $\cos i'$ term in $d\lambda'/dt$ can be transformed to the equatorial
coordinates by
\begin{equation}
\label{cosip}
	\cos i' = -\sin i \sin \lambda \sin i_t + \cos i \cos i_t.
\end{equation}

\section{Approximate Torques by the Inner Disk}
\label{appen:disk}

Here we estimate approximate torques on the outer disk by the inner
disk using the results obtained in Appendix \ref{appen:torus}.  If the
inner disk is considered in place of the torus, equations (\ref{coef20t})
and (\ref{coef21t}) can be written as
\begin{eqnarray}
\label{coef20d}
	J_{20}(r) & = & \frac{M_{id}}{2} \frac{R_{id}^2}{r^3}
                                                       (3 \sin^2 i_{id} - 2) \\
\label{coef21d}
	J_{21}(r) & = & \frac{M_{id}}{4} \frac{R_{id}^2}{r^3} \sin 2 i_{id},
\end{eqnarray}
and the rates of changes become
\begin{eqnarray}
\label{rate1d}
        \frac{di}{dt}       & \simeq & \frac{3G}{4r V_c}
                              2 J_{21} \sin (\lambda-\lambda_{id}) \cos i    \\
\label{rate2d}
	\frac{d\lambda}{dt} & \simeq & \frac{3G}{4r V_c} \left ( J_{20} \cos i
			      + 2 J_{21} \cos (\lambda - \lambda_{id})
                                \frac{\cos 2i}{\sin i} \right ),
\end{eqnarray}
where the variables with a subscript $id$ are for the inner disk and those
with no subscripts are for the outer disk.


\begin{deluxetable}{clcccclcc}
\tablecolumns{9}
\tablewidth{0pt}
\tablecaption{
\label{table:simul}Simulation Parameters}
\tablehead{
\colhead{} &
\colhead{} &
\colhead{} &
\multicolumn{4}{c}{Halo} &
\colhead{} &
\colhead{} \\ \cline{4-7}
\colhead{} &
\colhead{Model} &
\colhead{} &
\colhead{$a$} &
\colhead{$b$} &
\colhead{$c$} &
\colhead{Remark} &
\colhead{} &
\colhead{Torus}
}
\startdata
\sidehead{Initially Tilted Disks}
& TD-O    & & 10    & 10   & 7.5 & Oblate              & & No  \\
& TD-OL7  & & 10.7  & 9.3  & 7.5 & Oblate-like, 7~\%   & & No  \\
& TD-OL13 & & 11.3  & 8.7  & 7.5 & Oblate-like, 13~\%  & & No  \\
& TD-P    & & 7.5   & 7.5  & 10  & Prolate             & & No  \\
& TD-PL7  & & 8.03  & 6.98 & 10  & Prolate-like, 7~\%  & & No  \\
& TD-PL13 & & 8.48  & 6.53 & 10  & Prolate-like, 13~\% & & No  \\
\sidehead{Tilted Cosmic Infall}
& CI-S    & & 10    & 10   & 10  & Spherical           & & Yes \\
& CI-O    & & 10    & 10   & 7.5 & Oblate              & & Yes \\
& CI-OL7  & & 10.7  & 9.3  & 7.5 & Oblate-like, 7~\%   & & Yes \\
& CI-OL7b & & 9.3   & 10.7 & 7.5 & Oblate-like, 7~\%   & & Yes \\
& CI-P    & & 7.5   & 7.5  & 10  & Prolate             & & Yes \\
& CI-PL7  & & 8.03  & 6.98 & 10  & Prolate-like, 7~\%  & & Yes \\
& CI-PL7b & & 6.98  & 8.03 & 10  & Prolate-like, 7~\%  & & Yes \\
\enddata
\end{deluxetable}

\begin{deluxetable}{clcrrrcrrr}
\tablecolumns{10}
\tablewidth{0pt}
\tablecaption{
\label{table:jvalues}$J$ Values of Our Models}
\tablehead{
\multicolumn{3}{c}{} &
\multicolumn{3}{c}{$r=2$} &
\colhead{} &
\multicolumn{3}{c}{$r=4$} \\ \cline{4-6} \cline{8-10}
\colhead{} &
\colhead{Model} &
\colhead{} &
\colhead{$J_{20}$} &
\colhead{$J_{21}$} &
\colhead{$J_{22}$} &
\colhead{} &
\colhead{$J_{20}$} &
\colhead{$J_{21}$} &
\colhead{$J_{22}$}
}
\startdata
\sidehead{Halo}
&TD/CI-O    && $-$5.05E-2 &\nodata & \nodata && $-$6.67E-2 &\nodata & \nodata \\
&TD/CI-OL7  && $-$5.00E-2 &\nodata & 1.22E-2 && $-$6.62E-2 &\nodata & 1.62E-2 \\
&TD/CI-OL13 && $-$4.90E-2 &\nodata & 2.28E-2 && $-$6.49E-2 &\nodata & 3.09E-2 \\
&TD/CI-P    && 5.84E-2    &\nodata & \nodata && 7.61E-2    &\nodata & \nodata \\
&TD/CI-PL7  && 5.89E-2    &\nodata & 1.43E-2 && 7.66E-2    &\nodata & 1.81E-2 \\
&TD/CI-PL13 && 6.01E-2    &\nodata & 2.66E-2 && 7.78E-2    &\nodata & 3.36E-2 \\
\sidehead{Torus}
&All && $-$7.31E-3 & 1.02E-3 & $-$2.72E-4 && $-$2.93E-2 & 4.07E-3 & $-$1.09E-3
\enddata
\end{deluxetable}

\clearpage

\clearpage
\begin{references}
\reference{} Aarseth, S. J. \& Binney, J. 1978, MNRAS, 185, 227
\reference{} Allgood, B., Flores, R. A., Primack, J. R., Kravtsov, A. V.,
    Wechsler, R. H., Faltenbacher, A., \& Bullock, J. S. 2006, MNRAS, 367, 1781
\reference{} Ann, H. B. \& Park, J.-C. 2006, NewA, 11, 293
\reference{} Arnaboldi, M., \& Sparke, L. S. 1994, ApJ, 107, 958
\reference{} Bailin, J., \& Steinmetz, M. 2005, ApJ, 627, 647
\reference{} Bailin, J., et al. 2005, ApJ, 627, L17
\reference{} Bett,P., Eke, V., Frenk, C. S., Jenkins, A., Helly, J., \&
    Navarro, J. 2007, MNRAS, 376, 215
\reference{} Binney, J., Jiang, I.-G., \& Dutta, S. 1998, MNRAS, 297, 1237
\reference{} Binney, J., \& Tremaine, S. 2008, Galactic Dynamics, 2nd ed.,
    Sec. 2.5.3 (Princeton: Princeton University Press)
\reference{} Briggs, F. H. 1990, ApJ, 352, 15
\reference{} Chandrasekhar, S. 1969, Ellipsoidal Figures of Equilibrium
    (New York: Dover)
\reference{} Dehnen, W. 1993, MNRAS, 265, 250
\reference{} Dekel, A., \& Shlosman, I. 1983, in IAU Symp. 100, Internal
    Kinematics and Dynamics of Galaxies, ed. E. Athanassoula, (Dordrecht:
    Reidel Publishing Co.), 187
\reference{} Dubinski, J. 1992, ApJ, 401, 441
\reference{} Dubinski, J., \& Kuijken, K. 1995, ApJ, 442, 492
\reference{} Elwert, G., \& Hablick, D. 1965, ZA, 61, 273
\reference{} Habe, A., \& Ikeuchi, S. 1985, ApJ, 289, 540
\reference{} Hurnquist L. 1993, ApJS, 86, 389 
\reference{} Hunter, C., \& Toomre, A. 1969, ApJ, 155, 747
\reference{} Ideta, M., Hozumi, S., Tsuchiya, T., \& Takizawa, M. 2000, MNRAS,
    311, 733
\reference{} Jiang, I.-G., \& Binney, J. 1999, MNRAS, 303, L7
\reference{} Jing, Y. P., \& Suto, Y. 2002, ApJ, 574, 538
\reference{} Kahn, F. D., \& Woltjer, L. 1959, ApJ, 130, 705
\reference{} Kellogg, O. D. 1953, Foundations of Potential Theory (New York:
    Dover)
\reference{} Lopez-Corredoira, M., Betancort-Rijo, J., \& Beckman, J. E. 2002,
    A\&A, 386, 169
\reference{} Lynden-Bell, D. 1965, MNRAS, 129, 299
\reference{} Merritt, D., \& Fridman, T. 1996, ApJ, 460, 136
\reference{} Nelson, R. W., \& Tremaine, S. 1995, MNRAS, 275, 897
\reference{} Oh, S. H., Kim, W.-T., Lee, H. M., \& Kim, J. 2008, ApJ, 683, 94
\reference{} Ostriker, E. C., \& Binney, J. J. 1989, MNRAS, 237, 785
\reference{} Revaz, Y., \& Pfenniger, D. 2001, in Gas and Galaxy Evolution,
    ed. J. E. Hibbard, M. Rupen, \& J. H. van Gorkom, San Francisco, 
    ASP Conf. Proc., 240, 278
\reference{} Shen, Juntai \& Sellwood, J. A. 2006, MNRAS, 370, 2 
\reference{} Sparke, L. S., \& Casertano, S. 1988, MNRAS, 234, 873
\reference{} Springel, V., Yoshida, N.,\& White, S. D. M. 2001, NewA, 6, 79
\reference{} Springel, V., et al. 2005, Nature, 435, 629
\reference{} Steiman-Cameron, T. Y. \& Durisen, R. H. 1984, ApJ, 276, 101
\reference{} Toomre, A. 1983, in IAU Symp. 100, Internal Kinematics and
    Dynamics of Galaxies, ed. E. Athanassoula, (Dordrecht: Reidel Publishing
    Co.), 177
\reference{} Warren, M. S., Quinn, P. J., Salmon, J. K., \& Zurek, W. H.
    1992, ApJ, 399, 405
\end{references}
\end{document}